\documentclass[12pt]{article} 
\usepackage[pdftex]{graphicx}
\usepackage{epstopdf}
\usepackage{amsmath}
\usepackage{amssymb}
\usepackage{authblk}
\usepackage{mathrsfs}
\usepackage{slashed}
\usepackage{subcaption}

\makeatletter
 
  \@addtoreset{equation}{section}
 \makeatother

\newcommand{\be}{\begin{equation}}
\newcommand{\ee}{\end{equation}}
\newcommand{\bea}{\begin{eqnarray}}
\newcommand{\eea}{\end{eqnarray}}
\newcommand{\beann}{\begin{eqnarray*}}
\newcommand{\eeann}{\end{eqnarray*}}
\newcommand{\nn}{\nonumber}
\newcommand{\ba}{\begin{array}}
\newcommand{\ea}{\end{array}}

\DeclareMathOperator{\Tr}{Tr}
\DeclareMathOperator{\tr}{tr}

\DeclareMathOperator{\sdet}{Sdet}

\DeclareMathOperator{\Volume}{Vol}

\newcommand{\e}{\epsilon}

\newcommand{\zb}{{\bar{z}}}
\newcommand{\wb}{{\bar{w}}}
\newcommand{\A}{{\cal A}}
\newcommand{\del}{\partial}
\newcommand{\D}{\mathscr{D}}
\newcommand{\C}{\mathbb{C}}
\newcommand{\Z}{\mathbb{Z}}
\newcommand{\R}{\mathbb{R}}
\newcommand{\Nc}{{N_{\rm C}}}
\newcommand{\Nf}{{N_{\rm F}}}

\title{Higgs and Coulomb Branch Descriptions of the Volume of the Vortex Moduli Space}
\author[1]{Kazutoshi Ohta\thanks{kohta@law.meijigakuin.ac.jp}}
\author[2]{Norisuke Sakai\thanks{norisuke.sakai@gmail.com}}
\affil[1]{\small\it Institute of Physics, Meiji Gakuin University, Yokohama, Kanagawa 244-8539, Japan}
\affil[2]{\small\it Department of Physics, and Research and 
Education Center for Natural Sciences,
Keio University, 4-1-1 Hiyoshi, Yokohama, Kanagawa 223-8521, Japan, 
and 
iTHEMS, RIKEN, 
2-1 Hirasawa, Wako, Saitama 351-0198, Japan}
\date{}							

\begin{document}
\maketitle


\begin{center}
{\bf Abstract}
\end{center}
BPS vortex systems on closed Riemann surfaces with arbitrary 
genus are embedded into two-dimensional
supersymmetric Yang-Mills theory with matters.
We turn on
background $R$-gauge fields  
to keep half of rigid supersymmetry (topological A-twist) 
on the curved space. 
We consider two complementary descriptions; 
Higgs and Coulomb branches. 
The path integral reduces to the zero mode integral by the 
localization in the Higgs branch.
The integral over the bosonic zero modes directly gives an 
integral over the volume form of the moduli space, whereas 
the fermionic zero modes are compensated by an appropriate 
operator insertion.
In the Coulomb branch description with the same operator insertion, 
the path integral reduces to a finite-dimensional residue integral.
The operator insertion automatically determines
a choice of integral contours,
leading to the Jeffrey-Kirwan residue formula.
This result ensures the existence of the solution to the BPS vortex 
equation and explains the Bradlow bounds of the BPS vortex.
We also discuss a generating function of the volume of the 
vortex moduli space and show a reduction of the moduli space
from semi-local to local vortices.

\newpage

\section{Introduction}

Vortices play an important role in many physical phenomena 
in diverse area of physics, and give vital information on 
non-perturbative dynamics of gauge theories in two dimensions. 
When the quartic coupling of the Higgs scalar field is given by 
the square of the gauge coupling, static forces between vortices 
cancel, leaving vortex position and orientation as moduli 
parameters of the solution. These vortices are called 
BPS vortices \cite{Bogomolny:1975de, Prasad:1975kr}. 
In flat space, their characteristic features can 
be understood from symmetry, since the bosonic theory admitting 
BPS vortices can be embedded into supersymmetric theory and BPS 
vortices preserve half of supercharges \cite{Eto:2006pg}. 
Important generalizations of BPS vortices have been studied 
in curved space, such as 
hyperbolic space \cite{Witten:1976ck, Manton:2009ja, Manton:2010wu, 
Eto:2012aa} and general Riemann surfaces \cite{MantonSutcliffe}. 
The moduli space of BPS vortices allows interesting applications 
to many physical phenomena, including the thermodynamics of 
vortices \cite{Manton:1993tt, Eto:2007aw}. 
The volume of the moduli space is primarily obtained by an 
integration of the volume form, which is constructed from the 
metric, over the moduli space. 
It is generally difficult to construct an explicit metric 
for the moduli space of the vortices, except in simple situations 
such as well separated vortices \cite{Fujimori:2010fk}. 
However, it has been observed that the volume of the moduli 
space can be evaluated in the case of $U(1)$ gauge theory with 
a single flavor of charged scalar field, which is called the 
Abrikosov-Nielsen-Olesen (ANO) vortex, even though the metric 
for multi-vortices cannot be obtained explicitly \cite{Manton:1998kq}. 
One of the physically interesting properties of the moduli space 
of BPS vortices is the Bradlow bound: the BPS equations
admit solutions
only if the number of vortices is smaller than the area 
divided by the intrinsic size of BPS vortices \cite{Bradlow:1990ir}. 

In recent years, the localization method in supersymmetric field 
theories \cite{Pestun:2016zxk} has been developed and applied 
to evaluate various quantities exactly, including the partition 
function. 
In the localization method, it is essential to maintain 
some part of rigid supersymmetry on a curved manifold with 
isometry, such as the (squashed) three-sphere \cite{Hama:2010av, 
Kapustin:2009kz} and 
($\Omega$-deformed) two-sphere
\cite{Benini:2012ui,Doroud:2012xw,Closset:2013sxa,Closset:2015rna}. 
A few studies have also been done on 
the A-twist that may be applicable to general Riemann 
surfaces \cite{Witten:1988xj,Witten:1992xu,Closset:2014pda,Benini:2016hjo}. 
A systematic way of formulating rigid supersymmetry on curved space 
has been developed recently: twisting by a background $R$-gauge field 
plays a vital role and various choices generally give 
different types of rigid supersymmetry in curved space
\cite{Festuccia:2011ws, Dumitrescu:2012ha}. 
One should note that the usual choice of twist is applicable 
only for nice manifolds with isometry such as round or squashed sphere. 
On the other hand, we wish to consider vortices on arbitrary 
Riemann surfaces, that do not possess isometry in general. 
Moreover, the usual choice is not compatible with the BPS 
equation on the Riemann surface that we are looking for. 
In previous works, we have proposed a formalism to study 
the moduli space of vortices and other BPS solitons through 
the localization method, using a twisting different from the 
conventional one \cite{Moore:1997dj,Gerasimov:2006zt,Miyake:2011yr, Miyake:2011fq, Ohta:2013zna}. 
By inserting an appropriate operator, we can obtain the moduli 
space volume with this choice of twisting. We strengthen and develop 
our previously proposed method by studying two complementary 
descriptions and computing the effective action including fermionic 
terms explicitly in this work.

The purpose of this paper is to characterize the BPS vortex 
equations on Riemann surfaces $\Sigma_h$
 with a genus (handles) $h$ by embedding the theory into 
a rigid supersymmetric theory through a new choice of twisting 
by an $R$-gauge field background, and formulate the method to 
compute the moduli space volume of BPS vortices using the 
localization method.  
We here use a path integral formalism of $U(\Nc)$ gauge theory
with $\Nf$ flavors in the fundamental representation. 
To derive and understand the volume of the moduli space from 
the field theoretical (path integral) point of view, 
 the localization arguments in two different branches (phases) 
are important. 
To embed the BPS vortices, we will consider the supersymmetric 
Yang-Mills theory with vector and chiral multiplets. 
We also need to evaluate the vacuum expectation value (VEV) 
of an appropriate operator in order to obtain the volume of the 
vortex moduli space. 

If we consider localization of the path integral around the 
fixed point at non-zero values of the chiral multiplets, 
we find that the path integral gives an integral of the volume 
form over the moduli space. 
This is called the Higgs branch description of the volume of 
the moduli space.
The Higgs branch description is useful to demonstrate 
the physical meaning of moduli space volume directly. 
However, it is difficult to evaluate explicitly in general since 
we need an explicit metric to construct the volume form as we 
have mentioned.

Using the same field theory, we can evaluate the vacuum expectation 
value of the operator in an alternative Coulomb branch description. 
The localization method is so powerful in the Coulomb branch 
description that the path integral will reduce to a simple contour 
integral.
We can always perform this contour integral in principle. 
Since we are evaluating the same quantity in two different 
descriptions, this simple contour integral gives an alternative 
method of evaluating the moduli space volume. 
Although the relation between the contour integral (without knowing 
the metric explicitly) and the structure of the original moduli 
space is somewhat indirect, the field theory connects two 
different descriptions and explains why we can obtain the volume 
of the moduli space by the contour integral. 
To evaluate the path integral in this Coulomb branch description, 
we need to integrate over non-zero modes to find effective 
action for zero modes. 
Since we have inserted an operator, we also need to evaluate 
terms contributing to the correction to the operator, including 
fermionic terms. This point is also an improvement over our 
previous works \cite{Miyake:2011yr, Miyake:2011fq, Ohta:2013zna}. 

We also consider a generating function of the moduli space volume.
We can take a sum over vorticity $k$ ignoring the Bradlow bound
for an asymptotically large area $\A$.
The leading $\A$ behavior of the volume is found to reduce
from $\A^{k N}$ to $\A^k$
in the case of local vortices ($\Nc=\Nf= N$) on the sphere.
The generating function allows us to show the reduction
for arbitrary values of $N$ and $k$,
improving our previous result \cite{Miyake:2011yr}.

The organization of this paper is as follows. 
In Sect.\,\ref{sc:bps_eq}, 
we review the BPS equations on general Riemann surfaces. 
In Sect.\,\ref{sc:susy_curved}, we introduce ${\cal N}=(2,2)$ 
supersymmetric field theory and twisting by background $R$-gauge 
field to obtain the rigid supersymmetry. 
In Sect.\ref{sc:higgs}, the Higgs branch description is given, leading 
to the physical meaning as the volume of the vortex moduli space. 
In Sect.\ref{sc:coulomb}, the Coulomb branch description is given, leading 
to a simple contour integral formula. 
Section \ref{sc:generating_func} gives a generating function of 
the moduli space volume that leads to the reduction of
the moduli space in the case of the local vortex.
Section \ref{sc:discussion} is devoted to the conclusion and discussion. 
Appendix \ref{sc:cartan_weyl} gives the Cartan-Weyl basis and 
Appendix \ref{sc:heat_kernel} gives the 
heat kernel regularization.

\section{BPS Vortices 
 on Curved Riemann Surfaces}
\label{sc:bps_eq}

We first consider $2+1$ dimensional Yang-Mills-Higgs theory on 
$\R_t \times \Sigma_h$ to study 
a vortex system on the two-dimensional curved Riemann surface  $\Sigma_h$ 
with genus $h$. The space-time metric is given by 
\begin{equation}
ds_{ 2{+}1}^2=G_{MN}dx^M \otimes dx^N=-dt^2+g_{\mu\nu}dx^\mu \otimes dx^\nu,
\label{eq:metric}
\end{equation}
where $M, N=0,1,2$ and $\mu, \nu=1,2$.
We can take the metric of the Riemann surface to be conformally flat
\be
g_{\mu\nu}dx^\mu \otimes dx^\nu =  2 g_{z\zb} dz \otimes d\zb,
\nn
\ee
in suitable complex coordinates $z=x^1+ix^2$ and $\zb = x^1 - i x^2$
to be $g_{z\zb}=g_{\zb z}=\frac{1}{2}e^{2\rho(z,\zb)}$.

We are interested in $U(\Nc)$ gauge theory
with $\Nf$ flavors of scalar fields in the fundamental 
representation as an $\Nc \times \Nf$ matrix $H$. 
The action is given by 
\begin{equation}
 S = \int d^3x \sqrt{G} 
\Tr \left[-\frac{1}{2g^2}G^{MK} G^{NL} F_{MN} F_{KL}
- G^{MN}\D_M H \D_N H^\dagger -\frac{g^2}{4}
(\zeta {\bf 1}_\Nc-H H^\dagger)^2
\right], 
\label{eq:mdl:bosonic_action}
\end{equation}
where the covariant derivatives and field strengths are
defined as 
${
{\D}}_M H=(\partial_M + iA_M)H$, 
$F_{MN}=-i[{
{\D}}_M,\,{
{\D}}_N]$. 
It has been noticed that the Bogomolnyi completion can be 
found to give a topological bound for the energy $E$ of static 
configurations (in the $A_0=0$ gauge) 
\begin{eqnarray}
E&=& \int_{\Sigma_h} d^2 z \sqrt{g} \Tr \left[
-\frac{1}{g^2} (g^{z\zb}F_{z\zb})^2
+{
g^{z\zb}{\D}}_z H {
{\D}}_\zb H^\dagger
+{
g^{\zb z}{\D}}_\zb H {
{\D}}_z H^\dagger
+ \frac{g^2}{4}(\zeta {\bf 1}_\Nc-H H^\dagger)^2\right]
\nonumber \\
&=&\int_{\Sigma_h} d^2 z \sqrt{g}\Tr\left[\frac1{g^2}
\left(-ig^{z\zb}F_{z\zb} + {g^2 \over 2} 
(\zeta {\bf 1}_\Nc - H H^\dagger)\right)^2
+2g^{z\zb}\D_\zb H \D_z H^\dag 
+ i\zeta g^{z\zb}F_{z\zb}
\right]\nonumber \\
&\ge&
2\pi \zeta k , 
\label{eq:bogomolnyi_completion}
\end{eqnarray}
where we have dropped the total derivative term $2g^{z\bar z}\partial_{[z} H \D_{\bar z]} H^\dagger$
due to the compactness of the Riemann surface $\Sigma_h$.
The vorticity $k$
\begin{equation}
k=\frac{i}{2\pi}\int_{\Sigma_h} d^2 z \sqrt{g} \, g^{z\zb}F_{z\zb} \in \Z_{\geq 0},
\end{equation} 
measures the winding number of the $U(1)$ part of broken 
$U(N_{\rm C})$ gauge symmetry. 
The bound is saturated if and only if the following BPS 
equations are satisfied 
\begin{eqnarray}
&& -ig^{z\zb}F_{z\zb} +\frac{g^2}{2}(\zeta {\bf 1}_\Nc - HH^\dag) = 0,
\label{eq:vtx:BPSeq1} \\
&&\D_\zb H = \D_z H^\dag = 0 .
\label{eq:vtx:BPSeq2}
\end{eqnarray}

In the flat space, the above 
theory can be embedded in a supersymmetric theory, and the BPS 
vortices preserve precisely  half of supersymmetry. 
This is the reason why BPS vortices have many nice features 
such as no static force between vortices, resulting in vast 
moduli space for multi-vortices. 

Even in curved space-time such as $\R_t \times \Sigma_h$, 
the above BPS equations have been found and several interesting 
properties such as moduli space volume have been obtained in 
the case of $U(1)$ vortices \cite{MantonSutcliffe}. 
These vortices on Riemann surfaces enjoy many nice features 
similar to those on flat space. 
With the development of detailed understanding of supersymmetry 
on curved space-time \cite{Dumitrescu:2012ha}, it is now 
possible to understand more precisely the relation between 
BPS vortices on Riemann surfaces $\Sigma_{h}$ and the 
supersymmetry on curved space-time, which 
will be clarified in subsequent sections.

\section{Supersymmetric QCD on Curved Riemann Surfaces}
\label{sc:susy_curved}

The Bogomolnyi completion of the energy of the vortices on the curved Riemann surface
discussed in the previous section can be naturally embedded in 
two-dimensional ${\cal N}{=}(2,2)$ supersymmetric gauge theory.
The supersymmetric theory with four supercharges is obtained
by a dimensional reduction from four-dimensional ${\cal N}=1$ supersymmetric theory.
Apart from the $2+1$ dimensional Lorentzian metric of the Yang-Mills-Higgs theory
of the vortex system,
we first introduce four dimensional Euclidean space-time $\Sigma_h\times T^2$
to construct the rigid supersymmetry on $\Sigma_h$.

The metric of the four dimensional space-time is 
\be
ds_{4d}^2 = 2 g_{z\zb} dz \otimes d\zb + dw \otimes d\wb,
\nn
\ee
where
$w=x^3+ix^4,\wb=x^3-ix^4$ are complex coordinates on a flat two-torus $T^2$.
This is a particular choice among the more general metric discussed 
in Ref.~\cite{Dumitrescu:2012ha}.
After dimensional reduction along the torus $T^2$, we can define a rigid supersymmetry
with four supercharges on $\Sigma_h$.

We first consider the vector multiplet part.
The gauge fields in four dimensions reduce to  gauge field 
$A_\mu$ ($\mu=z,\zb$) and  adjoint scalar fields $X_i$ ($i=w,\wb$).
Spinors in four dimensions reduce to those in two dimensions. 
If we consider the spinors in space-time with the Euclidean 
signature, the two spinors in the vector multiplet, $\lambda$ 
and $\bar{\lambda}$ are independent of each other. 
On the other hand, $\bar{\lambda}$ is related to $\lambda$ 
by complex conjugation in the Lorentzian space-time. 
We will consider the Euclidean case, although the following 
arguments are also essentially valid in the Lorentzian signature. 
We will use the same notation as in \cite{Hama:2010av, Assel:2014paa} 
in the following. 

Following Refs.~\cite{Festuccia:2011ws,Dumitrescu:2012ha}, let us 
consider Killing equations for the supersymmetry transformation parameters
on the curved Riemann surface
\be
\begin{split}
\nabla^R_\mu \xi &\equiv  \nabla_\mu \xi + i \A^R_\mu \xi =0,\\
\nabla^R_\mu \bar{\xi} &\equiv  \nabla_\mu \bar{\xi} - i \A^R_\mu \bar{\xi} =0,
\end{split}
\label{covariantly constant}
\ee
where spin connections in the covariant derivative $\nabla_\mu$ are written in terms of the exponent of the conformal
factor $\rho(z,\zb)$ and 
$\A^R_\mu$ is a background gauge field
associated with a gauged $U(1)_R$ R-symmetry.
This equation appears as a transformation of the gravitino 
in the new minimal supergravity.
Note here that there is no solution to the 
eq.~(\ref{covariantly constant})
if $\A^R_\mu=0$, except for $h=0$.
As we will see, the eq.~(\ref{covariantly constant}) 
has a solution by choosing $\A^R_\mu$ suitably, since 
the background $U(1)_R$ gauge field can compensate 
the curvature of the Riemann surface.

The Lagrangian density
for the vector multiplet is given by
\be
{\cal L}_v = \frac{1}{g_0^2}\Tr\left\{
\frac{1}{2}F_{\mu\nu}F^{\mu\nu}
+\D_\mu X_i \D^\mu X^i
+2i\bar{\lambda}\bar{\sigma}^\mu \D^R_\mu \lambda
-\frac{1}{2}[X_i,X_j]^2
-2\bar{\lambda}\bar{\sigma}^i [X_i, \lambda]
-D^2
\right\},
\label{vector action}
\ee
where $g_0$ is a gauge coupling constant for the supersymmetric theory.
The field strength and covariant derivatives are defined by
\be
\begin{split}
F_{z\zb} &= \del_z A_\zb -\del_\zb A_z + i[A_z,A_\zb],\\
\D^R_\mu \lambda &= \nabla_\mu \lambda + i[A_\mu,\lambda] 
+ i \A^R_\mu \lambda,\\
\D^R_\mu \bar{\lambda} &= \nabla_\mu \bar{\lambda} 
+ i[A_\mu,\bar{\lambda}] - i \A^R_\mu \bar{\lambda},\\
\D_\mu X_i &= \nabla_\mu X_i + i[A_\mu,X_i].
\end{split}
\nn
\ee
The action (\ref{vector action}) is invariant under the supersymmetry 
transformation with 
$\delta = \xi^\alpha Q_\alpha+\bar{\xi}_{\dot{\alpha}}\bar{Q}^{\dot{\alpha}}$:
\be
\begin{split}
\delta A_\mu &= -i \bar{\lambda}\bar{\sigma}_\mu \xi + i \bar{\xi}\bar{\sigma}_\mu \lambda,\\
\delta X_i &= -i \bar{\lambda}\bar{\sigma}_{i} \xi+ i \bar{\xi}\bar{\sigma}_{i} \lambda,\\
\delta \lambda &= 2\sigma^{z\zb}\xi F_{z\zb}\
+2\sigma^{\mu i}\xi\D_\mu X_i+2i\sigma^{w\wb}\xi [X_w,X_\wb]+i\xi D,\\
\delta \bar{\lambda} &= -2\bar{\xi}\bar{\sigma}^{z\zb}F_{z\zb}
-2\bar{\xi}\bar{\sigma}^{\mu i}\D_\mu X_i
-2i\bar{\xi}\bar{\sigma}^{w\wb}[X_w,X_\wb]-i\bar{\xi}D,\\
\delta D &= -\xi \sigma^\mu \D^R_\mu \bar{\lambda}-\D^R_\mu\lambda \sigma^\mu \bar{\xi}
 -i [X_i, \xi \sigma^i\bar{\lambda}]-i[X_i,\lambda \sigma^i \bar{\xi}],
\end{split}
\nn
\ee
up to total derivatives since the supersymmetry transformation parameters 
$\xi$ and $\bar{\xi}$ are covariantly constant in the background 
$U(1)_R$ gauge field.

In order to preserve part of supersymmetry on the curved space, 
we have to find a solution
to (\ref{covariantly constant}). 
Fortunately, it is easy to find the solution to (\ref{covariantly constant})
in two dimensions. Indeed, if we set 
$\A^R_z=\frac{i}{2}\del_z \rho$ and 
$\A^R_\zb=-\frac{i}{2}\del_\zb \rho$,
they cancel some of the $\rho$ with the spin connections in the covariant derivatives and
the Killing equations (\ref{covariantly constant}) for each component of $\xi$ reduce to
\be
\begin{split}
\nabla^R_z \xi_1 &= (\del_z -\frac{1}{2}\del_z \rho + i \A_z^R)\xi_1 =( \del_z -\del_z \rho) \xi_1 =0,\\
\nabla^R_z \xi_2 &= (\del_z +\frac{1}{2}\del_z \rho + i \A_z^R)\xi_2 = \del_z \xi_2 =0,\\
\nabla^R_\zb \xi_1 &= (\del_\zb +\frac{1}{2}\del_\zb \rho + i \A_\zb^R)\xi_1 = (\del_\zb+\del_\zb \rho) \xi_1 =0,\\
\nabla^R_\zb \xi_2 &= (\del_\zb -\frac{1}{2}\del_\zb \rho + i \A_\zb^R)\xi_2 = \del_\zb \xi_2 =0,
\end{split}
\nn
\ee
and similarly for $\bar{\xi}$.
Then the equation (\ref{covariantly constant}) has a solution
\be
\xi_\alpha =\frac{1}{\sqrt{2}}
\begin{pmatrix}
0\\
\xi_0
\end{pmatrix},
\quad
\bar{\xi}^{\dot{\alpha}} = 
\frac{1}{\sqrt{2}}
\begin{pmatrix}
\bar{\xi}_0\\
0
\end{pmatrix},
\label{eq:const_spinor}
\ee
where $\xi_0$ and $\bar{\xi}_0$ are constant spinors.

By using explicit representation of the spinors and gamma matrices, 
we obtain the supersymmetry transformations in terms of components:
\be
\begin{split}
\delta A_z &= -\frac{i}{\sqrt{2}}\bar{\xi}_0 e^\rho \lambda_1,\\
\delta A_\zb &= -\frac{i}{\sqrt{2}} \xi_0 e^\rho \bar{\lambda}_{\dot{1}},\\
\delta X_w &= 0,\\
\delta X_{\bar{w}} &= \frac{i}{\sqrt{2}} (\xi_0 \bar{\lambda}_{\dot{2}} + \bar{\xi}_0 \lambda_2),\\
\delta \lambda_1 &= 2\sqrt{2}\xi_0 e^{-\rho}\D_z X_w,\\
\delta \lambda_2 &= \sqrt{2}\xi_0 e^{-2\rho} F_{z\zb}+i\sqrt{2}\xi_0 [X_w,X_\wb]+\frac{i}{\sqrt{2}}\xi_0 D,\\
\delta \bar{\lambda}_{\dot{1}} &= 2\sqrt{2}\bar{\xi}_0 e^{-\rho}\D_\zb X_w,\\
\delta \bar{\lambda}_{\dot{2}} &= -\sqrt{2}\bar{\xi}_0 e^{-2\rho} F_{z\zb}
+i\sqrt{2}\bar{\xi}_0 [X_w,X_\wb]-\frac{i}{\sqrt{2}}\bar{\xi}_0 D,\\
\delta D &= \sqrt{2}\xi_0 e^{-\rho} \D^R_z \bar{\lambda}_{\dot{1}}
-\sqrt{2}\bar{\xi}_0 e^{-\rho}\D^R_\zb \lambda_1
 -i\sqrt{2}\xi_0 [X_w, \bar{\lambda}_{\dot{2}}]
 +i\sqrt{2}\bar{\xi}_0 [X_w,\lambda_2].
\end{split}
\nn
\ee

Now let us choose to keep a single supercharge for a rigid 
supersymmetry $Q\equiv \frac{1}{\sqrt{2}}(Q_1+\bar{Q}_{\dot{1}})$, 
corresponding to the constant Grassmann parameter 
related by $\bar \xi_0=-\xi_0$ in eq.~(\ref{eq:const_spinor}). 
We obtain the following transformation properties of the fields 
\be
\begin{array}{lcl}
QA_\mu = \lambda_\mu, && Q \lambda_\mu = i\D_\mu \Phi,\\
Q\Phi =0, &&\\
Q\bar{\Phi} = 2\eta, && Q\eta = \frac{1}{2}[\Phi,\bar{\Phi}],\\
QD =
ig^{z\zb}(\D_z\lambda_\zb-\D_\zb\lambda_z)+[\Phi,\chi],
&& Q\chi = D-ig^{z\zb}F_{z\zb},
\end{array}
\nn
\ee
where we have also redefined the fields as follows:
\be
\begin{array}{lcl}
\Phi = 2X_w, && \bar{\Phi} = 2 X_\wb,\\
\lambda_z = \frac{i}{\sqrt{2}}e^\rho \lambda_1, && \lambda_\zb = -\frac{i}{\sqrt{2}}e^\rho \bar{\lambda}_{\dot{1}},\\
\eta = -\frac{i}{\sqrt{2}}(\lambda_2-\bar{\lambda}_{\dot{2}}), && \chi = -\frac{i}{\sqrt{2}}(\lambda_2+\bar{\lambda}_{\dot{2}}).
\end{array}
\nn
\ee
Note here that the dependence of the R-gauge field in the covariant derivative
$\D^R_\mu$ is absorbed into the factor $e^\rho$ in front of $\lambda_1$ and $\bar{\lambda}_{\dot{1}}$
so that $\lambda_z$ and $\lambda_\zb$ behave as 1-forms on $\Sigma_h$.

Furthermore if we define $Y\equiv D-ig^{z\zb}F_{z\zb}$, the algebra simply reduces to
\be
\begin{array}{lcl}
QA_\mu = \lambda_\mu, && Q \lambda_\mu = i\D_\mu \Phi,\\
Q\Phi =0, &&\\
Q\bar{\Phi} = 2\eta, && Q\eta = \frac{1}{2}[\Phi,\bar{\Phi}],\\
QY = [\Phi,\chi],
&& Q\chi = Y.
\end{array}
\ee
We find that the supercharge $Q$ is nilpotent up to a gauge 
transformation with the field $\Phi$ as the gauge parameter, 
\begin{equation}
Q^2 = -i\delta_{\Phi}.
\end{equation}

Using this $Q$ and the redefined fields, we can rewrite the 
Lagrangian for the vector multiplet in a $Q$-exact form
\be
{\cal L}_v = -\frac{1}{g_0^2} Q\Tr \left\{
ig^{z\zb}\lambda_z \D_\zb \bar{\Phi}
+ig^{\zb z} \lambda_\zb \D_z \bar{\Phi}
-\frac{1}{2}\eta [\Phi,\bar{\Phi}]
+\chi(Y- 2\mu_r)
\right\},
\ee
where $\mu_r =- ig^{z\zb}F_{z\zb}$ is a moment map associated 
with the D-term.

To turn on the FI term and the $\theta$-angle, we should add 
\be
{\cal L}_{\text{FI}+\theta} = \zeta Q\Tr \chi + i\tau 
\Tr g^{z\zb}F_{z\zb},
\ee
where $\tau=\zeta-i\theta$.
The first term can be absorbed into a redefinition of the moment map 
\be
\mu_r \to -ig^{z\zb}F_{z\zb} +\frac{g_0^2}{2} \zeta {\bf 1}_\Nc.
\ee

Next let us consider the chiral multiplet.
We have the $\Nf$ chiral multiplets (flavors) in the fundamental 
representation of the $U(\Nc)$ gauge group.
We represent them by an $\Nc \times \Nf$ matrix.
The chiral multiplet consists of boson $H$, fermion $\psi$ 
and auxiliary field $F$, which are assumed to have $U(1)_R$ 
charges $(r,r-1,r-2)$, respectively. 
By taking $\A^R_z=\frac{i}{2}\del_z \rho$ and 
$\A^R_\zb=-\frac{i}{2}\del_\zb \rho$, 
 the lowest component $H$ behaves as 
$(\frac{r}{2},0)$-form
on $\Sigma_h$.
The Lagrangian
\begin{multline}
{\cal L}_c = \Tr
\Big\{
\D_\mu^R H \D^{R\mu} H^\dag
+\frac{1}{2}|\bar{\Phi} H|^2
+\frac{1}{2} |\Phi H|^2
+ i\psi \sigma^\mu \D^R_\mu \bar{\psi}
+\psi \sigma^i \bar{\psi}X_i 
-F F^\dag\\
- i \sqrt{2}(\lambda\psi H^\dag  -  H \bar{\psi} \bar{\lambda})
-(D-\frac{r}{4}R)  H H^\dag
\Big\}
\label{eq:chiral_Lag}
\end{multline}
is invariant under the supersymmetry transformation
\be
\begin{split}
\delta H&= \sqrt{2}\xi^\alpha \psi_\alpha,\\
\delta \psi_\alpha &= i\sqrt{2}{\sigma^\mu}_{\alpha\dot{\alpha}} \bar{\xi}^{\dot{\alpha}} \D^R_\mu H
-\sqrt{2}{\sigma^i}_{\alpha\dot{\alpha}} \bar{\xi}^{\dot{\alpha}} X_i H + \sqrt{2} \xi_\alpha F,\\
\delta F &= i\sqrt{2} \bar{\xi}_{\dot{\alpha}} \bar{\sigma}^{\mu\dot{\alpha}\alpha} \D^R_\mu \psi_\alpha
-\sqrt{2} \bar{\xi}_{\dot{\alpha}} \bar{\sigma}^{i\dot{\alpha}\alpha} X_i \psi_\alpha + 2i\bar{\xi}_{\dot{\alpha}}\bar{\lambda}^{\dot{\alpha}} H,\\
\end{split}
\nn
\ee
where $R$ denotes the scalar curvature of the Riemann surface. 
Similar transformation properties apply for 
$H^\dagger, \bar \psi, F^\dagger$.
If we divide the two component fermions $\psi_\alpha$ into
 \be
\begin{array}{lcl}
\psi = \psi_1, &&\chi_\zb = e^\rho \psi_2,\\
\end{array}
\nn
\ee
the supersymmetry transformation becomes
\be
\begin{array}{lcl}
QH = \psi, && Q \psi = \Phi H,\\
QF=2ie^{-\rho}\D^R_\zb \psi+e^{-\rho}\Phi \chi_\zb+2i e^{-\rho}\lambda_\zb H,
&& Q \chi_\zb=e^\rho F -2i\D^R_\zb H,
\end{array}
\nn
\ee
in terms of the supercharge $Q$.

For later convenience, we define 
$Y_\zb \equiv e^\rho F -2i \D^R_\zb H$. 
Then the supersymmetry transformation in terms of $Q$ simply 
reduces to 
\be
\begin{array}{lcl}
QH = \psi, && Q \psi = \Phi H,\\
QY_\zb=\Phi \chi_\zb,
&& Q \chi_\zb= Y_\zb.
\end{array}
\ee
Similarly, for these complex conjugate fields, we have
\be
\begin{array}{lcl}
QH^\dag = \bar{\psi}, && Q \bar{\psi} =  -H^\dag \Phi,\\
QY_z= \chi_z \Phi,
&& Q \chi_z= -Y_z,
\end{array}
\ee
with $Y_z \equiv e^\rho F^\dag +2i \D^R_z H^\dag$.

Using this $Q$ and the redefined fields, we can write the Lagrangian 
as a $Q$-exact form
except for a part corresponding to the D-term
\be
{\cal L}_c = \frac{1}{2}Q \Tr \left\{
\psi H^\dag \bar{\Phi}
-\bar{\Phi} H \bar{\psi}
+\frac{1}{2}g^{\zb z} (Y_\zb -2\mu_\zb) \chi_z
-\frac{1}{2}g^{\zb z}  \chi_\zb (Y_z-2\mu_z)
\right\},
\nn
\ee
where $\mu_\zb= -2i\D^R_\zb H$ and $\mu_z=2i\D^R_z H^\dag$.
The term of $\Tr D H H^\dag$ can be absorbed into the moment map $\mu_r$
in the vector multiplet
by shifting
\be
\mu_r \to  -ig^{z\zb}F_{z\zb} + \frac{g_0^2}{2}(\zeta {\bf 1}_\Nc -HH^\dag).
\nn
\ee
This moment map consists of a part of the vortex equation.

To summarize, the total Lagrangian is written as a sum of the $Q$-exact 
part and $Q$-closed topological term
\be
{\cal L} ={\cal L}_0+i\tau\Tr g^{z\zb}F_{z\zb},
\label{Lagrangian density}
\ee
where
\be
{\cal L}_0 =  Q\left[\frac{1}{g_0^2} V_v +  V_c\right],
\nn
\ee
and
\bea
V_v &=& -\Tr \left\{
ig^{z\zb}\lambda_z \D_\zb \bar{\Phi}
+ig^{\zb z} \lambda_\zb \D_z \bar{\Phi}
-\frac{1}{2}\eta [\Phi,\bar{\Phi}]
+\chi(Y- 2\mu_r)
\right\},\\
V_c &=&  \frac{1}{2}\Tr \left\{
\psi H^\dag \bar{\Phi}
-\bar{\Phi} H \bar{\psi}
+\frac{1}{2}g^{\zb z} (Y_\zb -2\mu_\zb) \chi_z
-\frac{1}{2}g^{\zb z}\chi_\zb (Y_z-2\mu_z)
\right\},
\eea
with the moment maps
\bea
\mu_r &=& -ig^{z\zb}F_{z\zb} + \frac{g_0^2}{2}(\zeta {\bf 1}_\Nc -HH^\dag),
\label{real moment map}\\
\mu_\zb &=& -2i\D^R_\zb H,\\
\mu_z &=& 2i\D^R_z H^\dag.
\eea
The bosonic part of this Lagrangian reduces to the $2+1$ 
dimensional Yang-Mills-Higgs Lagrangian in eq.~(\ref{eq:mdl:bosonic_action}) 
after the auxiliary fields are eliminated. 
Note here that the auxiliary fields $Y$, $Y_z$ and $Y_\zb$ will 
give moment map constraints $\mu_r=\mu_\zb=\mu_z=0$, which are 
nothing but the BPS equations (\ref{eq:vtx:BPSeq1}) and 
(\ref{eq:vtx:BPSeq2}) of the vortices on the curved Riemann surface.
This is an essential reason why the supersymmetric theory gives 
the volume of the vortex moduli space (the space of solutions to the BPS
equations).

So far, we have assigned the generic R-charge for the components 
of the chiral multiplet $(H, \psi, \chi_\zb, Y_\zb)$. This means 
that we can consider a generalization of the 
vortex equations that contain higher spin (form) fields.
However our original vortex equations contain only the Higgs 
scalar (0-form) field $H$.
So we concentrate on a specific $R$-charge such that $r=0$ in the following.

\section{Higgs Branch Localization}
\label{sc:higgs}

\subsection{Coupling independence and fixed point equations}

Let us now consider a partition function for the supersymmetric 
theory that we have constructed in the previous section.
We will see that the partition function is closely related to the volume 
of the vortex moduli space.

The partition function is defined by the following path integral 
over the configuration space of the whole fields $\Psi$
\be
{\cal Z} = \int \frac{\D \Psi}{{\rm Vol}(G)}\, e^{-S[\Psi]},
\ee
where $\D \Psi$ is a suitable path integral measure for the 
fields and ${\rm Vol}(G)$ is a volume of the
gauge group $G=U(\Nc)$. The action $S[\Psi]$ is also written 
as a sum of the $Q$-exact and topological parts
through the Lagrangian density
(\ref{Lagrangian density})
\be
S[\Psi] = S_0[\Psi] +i\tau  \int_{\Sigma_h} d^2 z \sqrt{g} \, 
\Tr g^{z\zb}F_{z\zb}
\label{eq:sum_action}
\ee
where
\be
S_0[\Psi] =  Q\int_{\Sigma_h} d^2 z \sqrt{g} \left[
\frac{1}{g_0^2} V_v +  V_c
\right].
\label{Qexact-action}
\ee
Then the partition function is given by a summation over the 
possible topological sector 
\be
{\cal Z} = \sum_k {\cal Z}_k e^{-2\pi \tau k},
\label{field theory partition function}
\ee
where ${\cal Z}_k$ is defined by the path integral over the fixed 
topological sector with $k$ magnetic flux (vorticity) 
\be
{\cal Z}_k = \int \left[ \frac{\D \Psi}{{\rm Vol}(G)}\right]_k \, 
e^{-S_0[\Psi]}.
\nn
\ee

We will see that this ${\cal Z}_k$ is related to the volume of the 
moduli space of $k$ vortices.
So we concentrate only on ${\cal Z}_k$ for a while.

First of all, we note that ${\cal Z}_k$
is independent of the overall coupling of $S_0$.
For instance, if we consider the following partition function 
with rescaled action
\be
{\cal Z}_k(t) = \int  \left[ \frac{\D \Psi}{{\rm Vol}(G)}\right]_k \, 
e^{-tS_0[\Psi]},
\nn
\ee
we can show that a derivative of the partition function with 
respect to the parameter $t$
vanishes
\be
-\frac{\del}{\del t}{\cal Z}_k(t) = \left\langle Q\int_{\Sigma_h} 
d^2z \sqrt{g} \left[
\frac{1}{g_0^2} V_v +  V_c
\right] \right\rangle_k =0,
\ee
where $\langle \cdots \rangle_k$ stands for the vacuum expectation
value under the 
fixed topological sector, 
because of the $Q$-exactness of the action $S_0$.
Thus, we can conclude that the WKB (1-loop) approximation becomes 
exact in the limit of $t\to \infty$.
Similarly, we also find that the vacuum expectation value of 
the supersymmetric (cohomological)
operator ${\cal O}$, such that $Q{\cal O}=0$,
is not only independent of the parameter $t$ but also the gauge coupling $g_0$
in the $Q$-exact action.

After eliminating the auxiliary fields, the bosonic part of the 
$Q$-exact action $S_0$ becomes a sum of the positive definite pieces
\be
\left. S_0 \right|_B = \Tr\int_{\Sigma_h} d^2 z \sqrt{g}  \left[
\frac{1}{g_0^2}\left\{g^{\mu\nu}\D_\mu\Phi \D_\nu \bar{\Phi}
+\frac{1}{4}[\Phi,\bar{\Phi}]^2
+\mu_r^2
\right\}
+
\frac{1}{2}|\Phi H|^2 +\frac{1}{2} |\bar{\Phi}H|^2
+\frac{1}{2}g^{z\zb}\mu_z\mu_\zb
\right].
\nn
\ee
Using this coupling independence, we find that the path integral would be localized at solutions of a set of 
fixed (saddle) point equations as follows 
\bea
&&\D_\mu \Phi = 0,
\label{eq:fixed_point1}
\\
&&[\Phi,\bar{\Phi}] = 0,\\
&&\Phi H = \bar{\Phi} H = 0,\label{Phi-H}\\
&&\mu_r = \mu_z = \mu_\zb =0.
\label{moment map constrains}
\eea
We examine the solutions to the above fixed point equations later, but
we here focus on the eq.~(\ref{Phi-H}).
The eq.~(\ref{Phi-H}) gives two different kinds of the solution, i.e.,
\be
\Phi=0 \text{ and } H\neq 0,
\label{Higgs}
\ee 
or 
\be
\Phi\neq 0 \text{ and } H= 0.
\label{Coulomb}
\ee
We refer to each kind of  solution (\ref{Higgs}) and (\ref{Coulomb})
as the Higgs and Coulomb branch fixed points, respectively.

\begin{figure}
\begin{minipage}[b]{.5\linewidth}
\centering\includegraphics[scale=0.45]{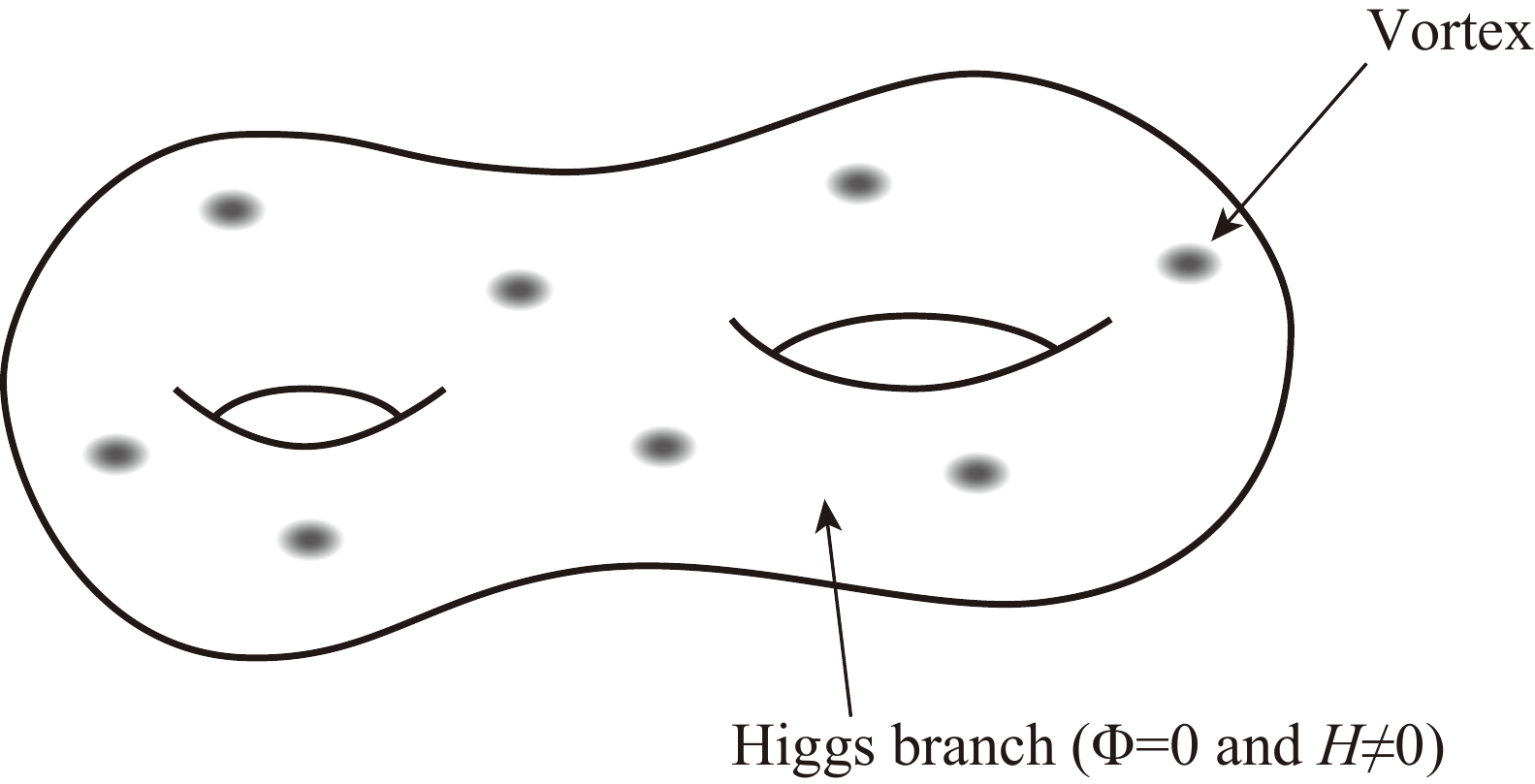}
\subcaption{{\footnotesize Small size vortices ($g_0^2\zeta \gg \frac{4\pi k}{\A}$)}}
\end{minipage}%
\begin{minipage}[b]{.5\linewidth}
\centering\includegraphics[scale=0.45]{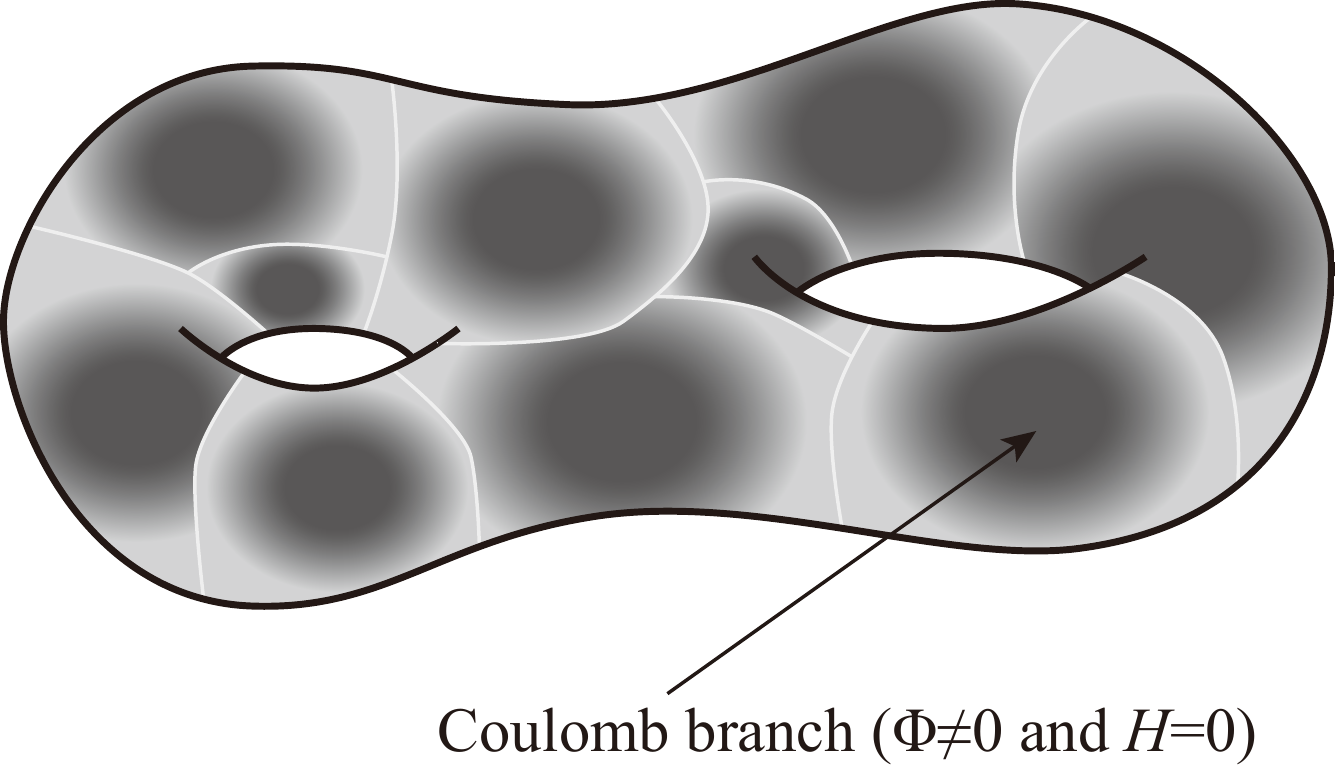}
\subcaption{{\footnotesize Vortices near the Bradlow bound 
($g_0^2\zeta \sim \frac{4\pi k}{\A}$)}}
\end{minipage}
\caption{
Two different descriptions of the vortices on the Riemann surface.
For the generic (small) size of the vortices 
($g_0^2\zeta \gg \frac{4\pi k}{\A}$), the Higgs vacuum dominates 
over the Riemann surface and the Higgs branch description is valid.
At the Bradlow bound ($g_0^2\zeta = \frac{4\pi k}{\A}$), the vortex 
size becomes large and the Coulomb branch inside the vortices 
occupies the whole Riemann surface. The Coulomb branch 
description is valid for this situation.
}
\label{Domains}
\end{figure}

The moment map constraints (the BPS vortex equations) in 
eq.~(\ref{moment map constrains}) say that $H$ does not vanish 
except at a finite number of points (vortex positions). 
These vortex solutions are compatible with the Higgs branch 
fixed points in eq.~(\ref{Higgs}), but incompatible with the Coulomb 
branch fixed points in eq.~(\ref{Coulomb}). 
In fact, the moment map constraint $\mu_r = 0$ with $H=0$ cannot 
be satisfied except for one particular value of the
couplings at $g_0^2\zeta = \frac{4\pi k}{\A}$
(the saturation point of the Bradlow bound). 
For the generic value of the couplings, we  need to consider the Higgs branch
description only and 
should not sum contributions from 
Higgs and Coulomb branch fixed points in performing the 
path integral (see Fig.~\ref{Domains} (a)).

Let us now elaborate on the possible significance of the 
Coulomb branch fixed points. 
From the exact solution of the ANO vortex ($\Nc=\Nf=1$), 
whose typical size is proportional to $1/g_0\sqrt{\zeta}$, we 
can see that the vev of the Higgs field inside the vortices
decreases rapidly. 
If we take the large size limit of the vortices $g_0^2 \zeta 
\to \frac{4\pi k}{\A}$ by 
making $g_0$ smaller, we encounter 
the upper bound of the vortex size due to the Bradlow bound 
($g_0^2 \zeta \geq \frac{4\pi k}{\A} $). 
When the Bradlow bound is saturated, 
the center cores of the vortices are enlarged and 
the Riemann surface can be filled up with the Coulomb vacua 
($\Phi\neq 0 \text{ and } H= 0$), where 
eq.~(\ref{moment map constrains}) is solved as 
\be
ig^{z\zb}F_{z\zb} = \frac{g_c^2\zeta}{2} \qquad
\text{and}\qquad
H = 0,
\ee
using a critical value $g_c$ for the coupling $g_0$, which 
is defined by $g_c^2 \zeta = \frac{4\pi k}{\A}$. 
In this situation, the magnetic flux uniformly spreads out on the Riemann 
surface. (See Fig.~\ref{Domains} (b).)

\begin{figure}
\centering\includegraphics[scale=1.0]{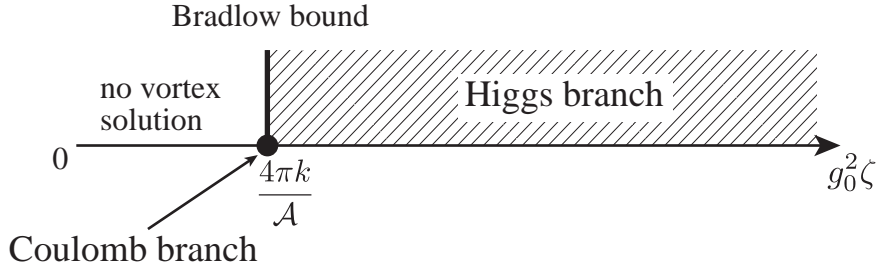}
\caption{
Complementary descriptions of the Higgs and Coulomb branches with respect to the gauge coupling and FI parameter
(vortex size) $g_0^2\zeta$.
In the general coupling region, the Higgs branch description is valid, but the Coulomb branch arises on the whole 
Riemann surface when the Bradlow bound is saturated.
Using the coupling independence of the localization theorem, the evaluations of the path integral are equivalent
to each other
and we can use the Coulomb branch description at the extreme couplings, instead of the Higgs branch localization.}
\label{descriptions}
\end{figure}

The localization theorem says that the path integral is 
independent of the coupling $g_0$. 
in the $Q$-exact action. 
Therefore we can tune the coupling $g_0$ to allow the 
Coulomb branch fixed point without changing the path integral results. 
Moreover, we can expect that the evaluation of the path integral 
(the partition functions or vevs of the cohomological operators) 
in the two different parameter regions gives the same 
answer\footnote{This gauge coupling $g_0$ in the $Q$-exact 
action can be different from the coupling $g$ in the vortex BPS 
equations which we are interested in. In section {4.4}, we will 
introduce other controllable coupling by inserting a cohomological 
operator.}, i.e., for the partition function, we obtain 
\be
{\cal Z}_k^{\text{Higgs}}(g_0 > g_c) 
= {\cal Z}_k^{\text{Coulomb}}(g_0=g_c)
.
\ee
(See Fig.~\ref{descriptions}.)


Thus, we can use the extreme Coulomb branch description of the path integral instead of the Higgs branch.
A similar complementarity of two descriptions between the Higgs and Coulomb branches
through the FI parameters is also 
discussed in the quiver quantum mechanics \cite{Denef:2002ru,Ohta:2015fpe}.

In the following, we first discuss the Higgs branch description, but we will see that it is difficult to evaluate the
path integral concretely in the Higgs branch. We will also see that the Coulomb branch description makes the evaluation
of the path integral easy. This equivalence of two different descriptions is our key point of the calculation of the volume of the vortex.

\subsection{Gauge fixing}

Since our model has $U(\Nc)$ gauge symmetry, we need to fix 
the gauge symmetry in the quantization. 
We adopt the Becchi, Rouet, Stora and Tyutin (BRST) formalism 
to fix the gauge symmetry.

Introducing the Faddeev-Popov (FP) ghosts $C$ and $\bar{C}$ and 
the Nakanishi-Lautrup (NL) field $B$, which are in the adjoint representation, 
we define the BRST transformations
\be
\begin{array}{lcl}
\delta_B C = iC^2,\\
\delta_B \bar{C} = 2iB, && \delta_B B = 0.
\end{array}
\nn
\ee
The BRST transformation acts on the fields as the gauge transformation
with replacing the gauge transformation parameter by $C$
\be
\begin{array}{l}
\delta_B A_\mu = -\D_\mu C,\\
\delta_B \Phi = i[C,\Phi],\\
\delta_B H = i C H,\\
\text{etc.}
\end{array}
\nn
\ee
Note that the BRST transformation is nilpotent $\delta_B^2=0$ as usual.

Once the gauge fixing function $f(A_\mu, \Phi,\bar{\Phi},H,H^\dag,B)$ is given,
a Lagrangian of the gauge fixing term and FP term
can be written in the $\delta_B$-exact form
\be
{\cal L}_{\text{GF+FP}} = \frac{i}{g_0^2}\delta_B \Tr \left(\bar{C} f\right).
\nn
\ee
The BRST symmetry of the above Lagrangian is apparent from the 
nilpotency of $\delta_B$, but this gauge fixing condition 
violates the supersymmetry. 
Similarly to the supersymmetry transformation in the Wess-Zumino 
gauge, this phenomenon suggests that we need to supplement the 
supersymmetry transformation by a compensating transformation 
associated with the gauge transformation in order to pull the 
field configuration back to the gauge fixing subspace. 
For that purpose, we consider a linear combination of the 
supercharge $Q$ and the BRST 
transformation $\delta_B$, as $Q_B\equiv Q + i\delta_B$ \cite{Pestun:2016zxk}. 
We find that the modified supercharge $Q_B$ becomes nilpotent, 
namely $Q_B^2=0$, provided that we make an additional assumption for 
the supersymmetry transformation of the ghost field 
\be
Q C = \Phi,
\nn
\ee
and $Q\bar{C}=QB=0$.

Using $Q_B$, we now introduce the total gauge fixed Lagrangian 
replacing the Lagrangian of $S_0$ in eq.~(\ref{eq:sum_action}) 
\be
\tilde{{\cal L}} \equiv Q_B\left[
\frac{1}{g_0^2}V_v + V_c + \frac{1}{g_0^2}\Tr \left(\bar{C} f\right)
\right],
\label{QB-exact action}
\ee
as a $Q_B$ exact form. 
Since $V_v$ and $V_c$ are gauge (BRST) and $Q$ invariant functions 
of the fields, this Lagrangian reduces to
\be
\begin{split}
\tilde{{\cal L}}
&= Q\left[
\frac{1}{g_0^2}V_v + V_c
\right]
+\frac{i}{g_0^2}\delta_B \Tr \left(\bar{C} f\right)
+\frac{1}{g_0^2}Q\Tr \left(\bar{C} f\right)\\
&={\cal L}_0 + {\cal L}_{\text{GF+FP}} +\frac{1}{g_0^2}
\Tr \left[\bar{C} (Q f)\right],
\label{gauge fixing terms}
\end{split}
\nn
\ee
using the definition of $Q_B$. 
The first and second terms are the ordinary gauge fixed Lagrangian 
in the BRST formalism. 
The extra last term $\Tr \left(\bar{C} Q f\right)$ can be 
neglected, since it can be absorbed by a shift of a field 
$\eta$ with a suitable choice of the gauge fixing function $f$, 
as we will see in the next subsection.

Since the total Lagrangian is written in the exact form of the 
nilpotent operator $Q_B$ 
and the measure is invariant under the $Q_B$-symmetry,
we can conclude that the path integral is invariant under the 
rescaling of the overall coupling 
\be
\tilde{\cal L} \to t \tilde{\cal L}.
\nn
\ee
Thus we can use the localization arguments again for the total 
gauge fixed Lagrangian. 
Hence we consider the localization for the $Q_B$-exact 
action instead of $Q$.

\subsection{Evaluation of the 1-loop determinant}

Now let us consider the 1-loop approximation of the $Q_B$-exact 
action (\ref{QB-exact action}).
The additional gauge fixing term in (\ref{QB-exact action}) 
imposes the gauge fixing condition, but the localization fixed 
points do not change from the original one in the $Q$-exact action.
In particular, the fixed points are given by solutions of the 
moment map constraints, i.e., the vortex equations
\be
\mu_r = \mu_z = \mu_\zb=0.
\nn
\ee

In the Higgs branch, we have fixed points of $\Phi=\bar{\Phi}=0$,
since the solution of the vortex equation gives $H\neq 0$ in general.
Once we obtain the classical solution, we expand the fields 
around the fixed points by
\be
\begin{split}
A_\mu &= \hat{A}_\mu + \frac{1}{\sqrt t}\tilde{A}_\mu,\\
H &= \hat{H} + \frac{1}{\sqrt t} \tilde{H},
\end{split}
\nn
\ee
where {\it hat} fields denote a classical solution, which 
satisfies the moment map constraints (vortex equations),
and {\it tilde} fields are fluctuations around them.
Similarly we also need to expand other fields around the zeros, but
it is just a rescaling of the fields like 
$\Phi\to \frac{1}{\sqrt{t}}\tilde{\Phi}$,
$Y\to \frac{1}{\sqrt{t}}\tilde{Y}$,
$\psi \to \frac{1}{\sqrt{t}}\tilde{\psi}$, etc.
We will omit the {\it tilde} for these rescaled fields including 
the FP ghosts and NL field for simplicity in the following.

Using this expansion (rescaling) of the fields, we also expand
the rescaled total Lagrangian up to the quadratic order of 
the fluctuations,
\begin{multline}
t {\cal L}_B =
\frac{1}{g_0^2}\Tr\Bigg[
\frac{1}{2}\Phi (-\hat{\D}^\mu\hat{\D}_\mu +  g_0^2\hat{H} \hat{H}^\dag) \bar{\Phi}
+\frac{1}{2}\bar{\Phi} (-\hat{\D}^\mu\hat{\D}_\mu +  g_0^2\hat{H} \hat{H}^\dag) \Phi \\
-Y^2
+2
Y \left\{
-ig^{z\zb}(\hat{\D}_z \tilde{A}_\zb-\hat{\D}_\zb \tilde{A}_z)
-\frac{g_0^2}{2}(\tilde{H}\hat{H}^\dag+\hat{H}\tilde{H}^\dag)
\right\} \\
-\frac{g_0^2}{2}g^{z\zb}Y_z Y_\zb
-ig_0^2g^{\zb z}\left\{
\hat{\D}_\zb \tilde{H}+i\tilde{A}_\zb \hat{H}
\right\}Y_z
+ig_0^2g^{\zb z}Y_\zb\left\{
\hat{\D}_z \tilde{H}^\dag -i \hat{H}^\dag \tilde{A}_z
\right\}
\Bigg]\\
+{\cal O}(1/\sqrt{t}),
\label{bosonic part}
\end{multline}
for the bosonic part and
\begin{multline}
t{\cal L}_F =\frac{1}{g_0^2}
\Tr\Bigg[
2ig^{z\zb} \lambda_z \hat{\D}_\zb \eta
+2i g^{\zb z} \lambda_\zb \hat{\D}_z \eta 
 -g_0^2 \psi \hat{H}^\dag \eta
 -g_0^2\eta \hat{H} \bar{\psi}\\
+\chi\left\{
2ig^{z\zb}(
\hat{\D}_z \lambda_\zb
-\hat{\D}_\zb \lambda_z)
+g_0^2(\psi\hat{H}^\dag
+\hat{H}\bar{\psi})
\right\}\\
+ig_0^2 g^{\zb z}\left\{
\hat{\D}_\zb \psi +i\lambda_\zb\hat{H}
\right\} \chi_z
-ig_0^2 g^{\zb z} \chi_\zb\left\{
\hat{\D}_z \bar{\psi}
-i\hat{H}^\dag \lambda_z
\right\}
\Bigg]\\
+{\cal O}(1/\sqrt{t}),
\end{multline}
for the fermionic part,
where $\hat{\D}_\mu$ means that the gauge field inside the covariant derivative is classical one.

Next let us consider the gauge fixing term.
To find a suitable gauge fixing function, we pay attention to the terms proportional to $\eta$:
\begin{multline}
\frac{2i}{g_0^2}\Tr \left\{
\eta\left(
g^{z\zb}(\hat{\D}_z \lambda_\zb
+\hat{\D}_\zb \lambda_z)
-i\frac{g_0^2}{2}(\psi \hat{H}^\dag - \hat{H}\bar{\psi})
\right)
\right\}\\
=\frac{2i}{g_0^2}\Tr \left\{
\eta Q\left(
\hat{\D}^\mu \tilde{A}_\mu
-i\frac{g_0^2}{2}(\tilde{H}\hat{H}^\dag -\hat{H} \tilde{H}^\dag)
\right)
\right\}.
\nn
\end{multline}
So if we adopt the gauge fixing function for the fluctuations by
\be
f(\tilde{A}_\mu,\tilde{H},\tilde{H}^\dag,B) = \hat{\D}^\mu \tilde{A}_\mu
-i\frac{g_0^2}{2}(\tilde{H}\hat{H}^\dag -\hat{H} \tilde{H}^\dag)
+\frac{1}{2}B,
\ee
the extra term $\frac{1}{g_0^2}\Tr\left[\bar{C}(Q f)\right]$ in the $Q_B$-exact gauge fixing Lagrangian (\ref{gauge fixing terms}) can be
absorbed by shifting
\be
\eta \to \eta + \frac{i}{2}\bar{C},
\nn
\ee
without changing the path integral, as expected.

Thus we obtain the gauge fixing and FP ghost Lagrangian for 
the above gauge fixing function
\be
\begin{split}
t{\cal L}_{\text{GF+FP}} &= 
\frac{i}{g_0^2}\delta_B \Tr(\bar{C}f)\\
&=\Tr\left[
-\frac{2}{g_0^2}Bf-
\frac{i}{2g_0^2}C
\left(
-\hat{\D}^\mu\hat{\D}_\mu
+g_0^2\hat{H}\hat{H}^\dag 
\right)\bar{C}
+\frac{i}{2g_0^2}\bar{C}
\left(
-\hat{\D}^\mu\hat{\D}_\mu
+g_0^2\hat{H}\hat{H}^\dag 
\right)C
\right],
\end{split}
\label{GF+FP}
\ee
where we have used the BRST transformation for the fluctuation
\be
\begin{split}
\delta_B \tilde{A}_\mu &= -\hat{\D}_\mu C,\\
\delta_B \tilde{H} &= i C \hat{H},
\end{split}
\nn
\ee
since the ghost $C$ is the same order as the fluctuations.

Comparing the bosonic part of the Lagrangian (\ref{bosonic part})
with the ghost kinetic term in (\ref{GF+FP}),
we immediately find that the 1-loop determinants for $\Phi$-$\bar{\Phi}$ and $C$-$\bar{C}$
are  canceled with each other completely.
Thus we can eliminate $\Phi$-$\bar{\Phi}$ and $C$-$\bar{C}$ from the Lagrangian.

For other fields, we now define sets of the bosonic and fermionic fields
by
\be
\begin{array}{lcl}
\vec{\cal B} = (\tilde{A}_\zb,\tilde{H}),
&&
\vec{\cal Y} = (Y+iB,Y_\zb),
\\
\vec{\cal B}^\dag = (\tilde{A}_z,\tilde{H}^\dag)^T,
&&
\vec{\cal Y}^\dag = (Y-iB,Y_z)^T,
\\
\vec{\cal F} = (\lambda_\zb,\psi),
&&
\vec{\cal X} = (\eta-\chi,\chi_\zb),\\
\vec{\cal F}^\dag = (\lambda_z,\bar{\psi})^T,
&&
\vec{\cal X}^\dag = (\eta +\chi,\chi_z)^T,
\end{array}
\nn
\ee
then the quadratic part of the Lagrangian is written simply as
\be
\begin{split}
t{\cal L}_B &= \Tr\left[
-\frac{1}{g_0^2}|\vec{\cal Y}|^2
+\vec{\cal Y} \hat{D}\vec{\cal B}^\dag
+\vec{\cal B}\hat{D}^\dag\vec{\cal Y}^\dag
\right],\\
t{\cal L}_F &= \Tr\left[
\vec{\cal X}  \hat{D}\vec{\cal F}^\dag
+\vec{\cal F}  \hat{D}^\dag \vec{\cal X}^\dag
\right],
\end{split}
\nn
\ee
where
\be
|\vec{\cal Y}|^2 = Y^2 + B^2 + \frac{g_0^2}{2}g^{z\zb}Y_z Y_\zb,
\nn
\ee
and
\be
\hat{D} \equiv
\begin{pmatrix}
\frac{2i}{g_0^2} g^{z\zb}\hat{\D}_\zb & -\hat{H}\\
g^{z\zb}\hat{H}^\dag & ig^{z\zb}\hat{\D}_z
\end{pmatrix},
\quad
\hat{D}^\dag \equiv
\begin{pmatrix}
\frac{2i}{g_0^2} g^{z\zb}\hat{\D}_z & g^{z\zb}\hat{H}\\
-\hat{H}^\dag & ig^{z\zb}\hat{\D}_\zb
\end{pmatrix}.
\nn
\ee


The 1-loop determinants of non-zero modes of the bosons and fermions 
cancel each other completely:
\be
\text{(1-loop det)} = \frac{\det'\hat{D}^\dag\hat{D}}
{\det'\hat{D}^\dag\hat{D}}=1,
\nn
\ee
where {\it prime} stands for omitting the zero modes (eigenvalues).

The bosonic zero modes are given by solutions to the linear equations
\be
\hat{D} \vec{\cal B}^\dag = 0,
\nn
\ee
i.e., $\ker \hat{D}$. Since this equation is a linearized vortex 
equation and
$\ker \hat{D}^\dag = \emptyset$ under our 
choice of the BPS vortex solution $\hat{A}_\mu$ and $\hat{H}$
we find that
\be
\dim {\cal M}_k = \dim \ker \hat{D}.
\label{dimension of the kernel}
\ee
On the other hand, the fermionic zero modes are given by the equations
\be
\hat{D}\vec{\cal F}^\dag = \hat{D}^\dag \vec{\cal X}^\dag = 0.
\nn
\ee
As we discussed for the bosonic zero modes, we have seen 
$\ker \hat{D}^\dag = \emptyset$ in the BPS vortex background.
So we can conclude that there is no zero mode in $\vec{\cal X}^\dag$,
and the number of zero modes in $\vec{\cal F}^\dag$ is the 
same as the number of bosonic zero modes, which is $\dim \ker \hat{D}$.

We need to integrate these bosonic and fermionic zero modes 
after integration of the non-zero modes, which gives the cancellation of 
the 1-loop determinant.

\subsection{Volume of the moduli space}

We have seen that the partition function of the fixed topological 
sector ${\cal Z}_k$ of our model itself
vanishes in general due to the existence of the fermionic zero modes.
As we discussed above, we expect that there exist the fermionic 
zero modes only in the fields $\vec{\cal F}_1$ and $\vec{\cal F}_2$,
i.e., $\lambda_\mu$, $\psi$ and $\bar{\psi}$.

In order to obtain a meaningful quantity from ${\cal Z}_k$, we 
need to insert some operator within the path integral,
which compensates the fermionic zero modes. However an arbitrary 
operator cannot be inserted since it spoils the localization 
arguments above. As mentioned before, the supersymmetric operator 
does not break the coupling independence, but
if we want a non-trivial (non-vanishing) quantity, we have to insert 
a $Q$-closed but not $Q$-exact operator ($Q$-cohomological operator).

$Q$-cohomological operators are classified in terms of the 
descent equations \cite{Witten:1990bs,Witten:1992xu}
\be
\begin{split}
Q {\cal O}_0 &= 0,\\
Q {\cal O}_1 &= d {\cal O}_0,\\
Q {\cal O}_2 &= d {\cal O}_1,
\end{split}
\label{descent equation}
\ee
whose $n$-form operators ${\cal O}_n$ are given by
\be
\begin{split}
{\cal O}_0 &= \Tr W(\Phi),\\
{\cal O}_1 &= -i\Tr W'(\Phi)  \lambda,\\
{\cal O}_2 &= \Tr  \left\{ -iW'(\Phi)F + \frac{1}{2}W''(\Phi)
\lambda \wedge \lambda\right\},
\end{split}
\ee
where $W(\Phi)$ is a polynomial of $\Phi$, the one-form $\lambda \equiv 
\lambda_z dz+ \lambda_\zb d\zb$, and the two-form\footnote{The 
two-form $F$ should not be confused with the auxiliary field 
of the chiral multiplet in eq.~(\ref{eq:chiral_Lag}).} 
$F\equiv 
dA+iA\wedge A$.

From the descent equation (\ref{descent equation}), we find that 
the integration of ${\cal O}_2$
\be
{\cal I}_2 = \int_{\Sigma_h} {\cal O}_2 
\nn
\ee
is $Q$-closed but not $Q$-exact, since $\Sigma_h$ is the compact 
Riemann surface.
If we insert the exponential of this $Q$-closed operator 
$e^{{\cal I}_2}$, the zero modes in $\lambda$ are compensated 
at least because of the bi-linear term of $\lambda$ in ${\cal I}_2$.
However this operator depends on the vacuum expectation value 
of $\Phi$ in general
and changes the value of ${\cal Z}_k$ excluding the zero modes.
This is undesirable for our purpose.

In order not to yield the extra contribution from the inserted 
operator, we need to modify $e^{{\cal I}_2}$ 
by adding other $Q$-closed terms to be
\begin{multline}
e^{\beta{\cal I}_V(g)}\\
 = \exp\left[\beta \Tr\int_{\Sigma_h} d^2 z 
\sqrt{g}\left\{
\Phi
\left(-ig^{z\zb} F_{z\zb}+\frac{g^2}{2}(\zeta {\bf 1}_{\Nc} - HH^\dag)\right)
-g^{z\zb}\lambda_z\lambda_\zb
-\frac{g^2}{2} \psi\bar{\psi}
\right\}
\right],
\end{multline}
where 
$g$ is an additional coupling constant that can differ from the 
coupling $g_0$ in the $Q$-exact action $S_0$ in eq.~(\ref{Qexact-action}). 
The parameter $\beta$ serves to count the dimension of the moduli space volume 
(the number of continuous moduli parameters). 
Note here that the vev of $e^{\beta{\cal I}_V(g)}$ explicitly 
depends on the parameter $\beta$ and the coupling $g$ (and also $\zeta$),
since the above operator is $Q$-closed but not $Q$-exact 
($Q$-cohomological), in contrast to the coupling $g_0$ in the 
$Q$-exact action. 
We identify this coupling $g$ in the inserted operator 
$e^{\beta{\cal I}_V(g,\zeta)}$ as the physical coupling for the BPS 
vortices that we study. 

According to the localization theorem, the vev of the $Q$-cohomological 
operator can be evaluated 
by the solutions to the fixed point equations.
As we explained before, we can evaluate the vev of the operator 
at any value of the gauge coupling $g_0$ in the $Q$-exact action 
 $S_0$ in eq.~(\ref{Qexact-action}), 
without changing the value.
Evaluating the path integral in the Higgs branch, we can choose 
the coupling $g_0$ in $S_0$ identical to $g$ in the inserted 
operator $e^{\beta{\cal I}_V}$. Then the moment map constraint 
(fixed point equation) $\mu_r$ in eq.~(\ref{real moment map}) becomes
\be
\mu_r = -ig^{z\zb}F_{z\zb} + \frac{g}{2}(\zeta {\bf 1}_\Nc -HH^\dag)=0.
\label{real moment map 2}
\ee
Since the solution to eq.~(\ref{real moment map 2}) eliminates 
the factor of $\Phi$ in the exponent of $e^{\beta{\cal I}_V}$,
the operator $e^{\beta{\cal I}_V}$  at $g_0=g$ in the Higgs branch fixed points
reduces to
\be
\left. e^{\beta{\cal I}_V(g)} \right|_{\text{Higgs branch fixed point}}= 
\exp\left[-\beta \Tr\int_{\Sigma_h} d^2 z 
\sqrt{g}\left\{
g^{z\zb}\lambda_z\lambda_\zb
+\frac{g^2}{2} \psi\bar{\psi}
\right\}
\right].
\ee
The bosonic part of this operator value at the Higgs branch fixed point
gives just unity, but the fermionic part
compensates all the fermionic zero modes
since the exponent contains bi-linear terms
of fermion pairs; $(\lambda_z,\lambda_\zb)$ and $(\psi,\bar{\psi})$.
After integrating over the fermionic zero modes,
only appropriate product of the fermionic pairs survives
to give a power of $\beta$ with a unit coefficient.
Hence the power of $\beta$ is given by a
sum of the number of fermionic zero modes,
namely the dimension of the moduli space
because of (\ref{dimension of the kernel}).

Since the operator $e^{\beta{\cal I}_V(g)}$ does not change the
bosonic part of the path integral at the Higgs branch coupling $g_0=g$, 
the path integral at the Higgs 
branch reduces to
the integral over the classical solution of the vortex equation
\be
\begin{split}
\left\langle
e^{\beta{\cal I}_V(g)}
\right\rangle^{g_0=g}_k
&={\cal N}\beta^{\dim_\C {\cal M}_k}\int  \D \hat{A}_\mu \D 
\hat{H} \D \hat{H}^\dag,
\end{split}
\nn
\ee
where
$\langle \cdots \rangle^{g_0=g}$ stands for the evaluation of the path integral
by using the $Q$-exact action with the same coupling $g_0=g$ as in the operator $e^{\beta{\cal I}_V(g)}$,
$\dim_\C {\cal M}_k$ is a complex dimension of the moduli 
space of $k$ vortices,
and ${\cal N}$ is a numerical constant that is associated with 
the normalization of the path integral measure.

Let us now rewrite the above integral in the field configuration 
space in terms of the moduli parameters, which 
parametrize the BPS vortex solution.
We denote the moduli parameters by complex coordinates $m^a$, 
which span the K\"ahler moduli space.
Changing the integral measure from the fields of $\hat{A}_\mu$, 
$\hat{H}$ and $\hat{H}^\dag$,
which are defined in the flat configuration space,
to the moduli parameters $m^a$, the Jacobian factor $\det G_{a\bar{a}}(m)$ 
will appear,
where $G_{a\bar{a}}$ is the K\"ahler metric of the moduli space.
So we obtain
\be
\begin{split}
\left\langle e^{\beta{\cal I}_V(g)}\right\rangle_k^{g_0=g}&=
{\cal N}\beta^{\dim_\C {\cal M}_k}
\int_{{\cal M}_k} \prod_{a=1}^{\dim_\C {\cal M}_k}
dm^a d\bar{m}^{\bar{a}}\det G_{a\bar{a}}(m)\\
&={\cal N}\beta^{\dim_\C {\cal M}_k}
\Volume {\cal M}_k,
\end{split}
\ee
where $\Volume {\cal M}_k$ is the volume of the $k$-vortex moduli space as we expected.

Thus we find that the path integral with the operator $e^{\beta{\cal I}_V(g)}$ insertion gives the volume of the moduli space.
However, to evaluate the above integral, we need to know the detail of the K\"ahler metric $G_{a\bar{a}}$,
but this is difficult in general.
We see that the $\langle e^{\beta{\cal I}_V(g)} \rangle_k^{g_0=g}$ is proportional to the volume of the moduli space in the Higgs branch description,
but we need to move into the Coulomb branch description to evaluate the volume explicitly.

Thanks to the localization theorem, we can also evaluate the 
above vev of the operator in the other coupling region without 
changing the value of the path integral, 
and can reach even extreme Coulomb branch couplings $g_0=g_c$, 
which satisfy $g_c^2\zeta=\frac{4\pi k}{\A}$.
Thus we can expect equivalence between the Higgs and Coulomb 
branch descriptions
\be
\left\langle e^{\beta{\cal I}_V(g)}\right\rangle^{g_0=g}_k 
=\left\langle e^{\beta{\cal I}_V(g)}\right\rangle^{g_0=g_c}_k,
\ee
using the same cohomological operator $e^{\beta{\cal I}_V(g)}$,
which measures the volume of the moduli space at the 
physical coupling $g$ 
for our BPS vortices. 
We emphasize that the evaluation of Coulomb branch path integral 
can be done using the $Q$-exact action $S_0$ with the critical 
coupling $g_c$, which differs from the physical coupling $g$ 
in the inserted operator $e^{\beta{\cal I}_V(g)}$. 


\section{Coulomb Branch Localization}
\label{sc:coulomb}

In this section, we consider the localization at the Coulomb branch,
where the fields are expanded around the fixed point solution 
with non-vanishing $\Phi$. 
In the following, we evaluate the path integral using the 
$Q$-exact action $S_0$ in eq.~(\ref{Qexact-action}) with the 
critical coupling $g_c$ defined by 
\be
g_c^2\zeta=\frac{4\pi k}{\A},
\label{critical_coupling}
\ee
which is different from the physical value of the coupling 
$g$ in the inserted operator $e^{\beta{\cal I}_V(g)}$.

We will discuss the general non-Abelian case, but to see an 
essence of the Coulomb branch localization,
we first explain the Abelian case.

\subsection{Abelian case}

In the Abelian theory, we denote the neutral scalar field by 
a lowercase letter $\phi$.
The fixed point equations $\del_\mu \phi=0$ in eq.~(\ref{eq:fixed_point1}) 
and $\phi H =0$ in eq.~(\ref{Phi-H}) say that $H$ 
vanishes if $\phi$ is a non-vanishing constant. 
We denote the solution to this fixed point equation by $\phi_0$ 
(constant zero mode).
The classical solution of the Abelian gauge field $a_\mu$ satisfies
\be
\frac{i}{2\pi}\int_{\Sigma_h} d^2 z (\del_z a_\zb - \del_\zb a_z) 
= k \in {\mathbb Z}_{\geq 0},
\nn
\ee
which is fixed while integrating the fluctuations in the $k$-vortex sector.

We now expand the bosonic fields in the vector multiplet around 
the classical solution (fixed points) by
\be
\begin{split}
\phi &= \phi_0 + \frac{1}{\sqrt{t}} \tilde{\phi},\\
A_\mu &= a_\mu + \frac{1}{\sqrt{t}} \tilde{A}_\mu,
\end{split}
\nn
\ee
and the auxiliary field $Y$ is also rescaled by 
$Y \to \frac{1}{\sqrt{t}} \tilde{Y}$.

For the fermionic fields, we expect that there are two 0-form 
zero modes and $2h$ 1-form zero modes on the Riemann surface 
with the genus $h$, since these zero modes are associated with 
0th and 1st cohomology on $\Sigma_h$, respectively.
We denote the 0-form zero modes by $\eta_0$, $\chi_0$ and 1-form 
zero modes by $\lambda_{0,z}
=\sum_{l=1}^h \lambda^{(l)}_{0}\omega^{(l)}$ and $\lambda_{0,\zb}
=\sum_{l=1}^h \bar{\lambda}^{(l)}_{0}\bar{\omega}^{(l)}$, where 
$\omega^{(l)}$ and $\bar{\omega}^{(l)}$ take values in the bases 
of $H^{1,0}(\Sigma_h,\Z)$ and $H^{0,1}(\Sigma_h,\Z)$, respectively. 
The 1-form bases are normalized by
\be
\int_{\Sigma_h} d^2 z \, \omega^{(l)}\bar{\omega}^{(l')}= \delta_{ll'}.
\ee

Thus we also expand the fermionic fields in the vector multiplets 
around these zero modes as
\be
\begin{split}
\eta &= \eta_0 + \frac{1}{\sqrt{t}} \tilde{\eta},\\
\chi &= \chi_0 + \frac{1}{\sqrt{t}} \tilde{\chi},\\
\lambda_\mu &=  \lambda_{0,\mu} + \frac{1}{\sqrt{t}} \tilde{\lambda}_\mu.
\end{split}
\nn
\ee

In contrast to the Higgs branch evaluation, the fixed point 
solution of the bosonic field $H$ vanishes.
So we rescale the bosons and fermions in the chiral multiplet as
\be
H \to \frac{1}{\sqrt{t}} \tilde{H},\quad
Y_\mu \to \frac{1}{\sqrt{t}} \tilde{Y}_\mu,\quad
\psi \to \frac{1}{\sqrt{t}} \tilde{\psi},\quad
\chi_\mu = \frac{1}{\sqrt{t}} \tilde{\chi}_\mu.
\nn
\ee
This rescaling is always guaranteed by the invariance of the 
path integral measure
\be
\D^2 H \D^2 Y_\mu \D^2 \psi \D^2 \chi_\mu.
\nn
\ee

Using the above expansion and rescaling, we find that the 
Lagrangian becomes just quadratic order in the fluctuations
\be
\begin{split}
t{\cal L} &= 
\frac{1}{g_c^2}\del_\mu \tilde{\phi} \del^\mu \bar{\tilde{\phi}}
- \frac{1}{g_c^2} \tilde{Y} ( \tilde{Y} +2 i g^{z\zb}(\del_z \tilde{A}_\zb - \del_\zb \tilde{A}_z) )\\
&\qquad + 2i \eta g^{z\zb}(\del_z \tilde{\lambda}_\zb + \del_\zb \tilde{\lambda}_z)
- 2i \chi g^{z\zb}(\del_z \tilde{\lambda}_\zb - \del_\zb \tilde{\lambda}_z)\\
&\qquad\qquad +\vec{\cal V} {\cal M} \vec{\cal V}^\dag +{\cal O}(1/\sqrt{t}),
\end{split}
\nn
\ee
where $\vec{\cal V} = (\tilde{H},\tilde{\psi},\tilde{\chi}_\zb)$,
$\vec{\cal V}^\dag = (\tilde{H}^\dag,\tilde{\bar{\psi}},
\tilde{\chi}_z)^T$ and
\be
{\cal M} = 
\begin{pmatrix}
-2g^{z\zb}\hat{\D}^{(1)}_\zb\hat{\D}^{(0)}_z + |\phi_0|^2 & -(\eta_0+\chi_0)
& -g^{z\zb}\lambda_{0,\zb}\\
-(\eta_0-\chi_0) & -\bar{\phi}_0 & -ig^{z\zb}\hat{\D}^{(1)}_\zb\\
-g^{z\zb}\lambda_{0,z} & -ig^{z\zb}\hat{\D}^{(0)}_z & \frac{1}{2}g^{z\zb}\phi_0
\end{pmatrix}.
\nn
\ee
The covariant derivatives $\hat{\D}^{(n)}_\mu$ are acting on the 
$n$-form field.

If the Lorentz gauge $\del_\mu \tilde{A}^\mu = 0$ is chosen,
the gauge fixing and FP terms are given by
\be
t{\cal L}_{\rm GF+FP} = \frac{i}{g_c^2} c(-\del_\mu\del^\mu) \bar{c}
-\frac{1}{g_c^2}B(B+2g^{z\zb}(\del_z \tilde{A}_\zb + \del_\zb \tilde{A}_z)).
\ee
Combining the quadratic part of the Lagrangian and the gauge fixing and FP terms,
1-loop determinants from 
the bosonic fields $(\tilde{\phi},\bar{\tilde{\phi}},\tilde{Y},B)$ and
the fermionic fields $(c,\bar{c},\tilde{\eta},\tilde{\chi})$
completely give the same contribution and
cancel each other (just giving one).

On the other hand, the 1-loop determinant from the chiral multiplets 
is given by
\be
\frac{1}{\sdet {\cal M}},
\nn
\ee
after integrating out the fluctuations $\vec{\cal V}$ and $\vec{\cal V}^\dag$.
If we use a decomposition of ${\cal M}$ by
\be
{\cal M} = \begin{pmatrix}
A & B\\
C & D
\end{pmatrix},
\nn
\ee
where
\be
\begin{split}
A &= -2g^{z\zb}\hat{\D}^{(1)}_\zb\hat{\D}^{(0)}_z + |\phi_0|^2,\\
B &= \begin{pmatrix}-(\eta_0+\chi_0)& -g^{z\zb}\lambda_{0,\zb}\end{pmatrix},\\
C &= \begin{pmatrix}-(\eta_0-\chi_0)\\-g^{z\zb}\lambda_{0,z}\end{pmatrix},\\
D &= \begin{pmatrix} -\bar{\phi}_0 & -ig^{z\zb}\hat{\D}^{(1)}_\zb\\-ig^{z\zb}\hat{\D}^{(0)}_z & \frac{1}{2}g^{z\zb}\phi_0\end{pmatrix},
\end{split}
\nn
\ee
then the superdeterminant is given by the determinants of the decompositions
\be
\frac{1}{\sdet{\cal M}} = \frac{\det D}{\det A}\det(1- X) 
= \frac{\det D}{\det A}e^{\Tr\log(1- X)},
\nn
\ee
where $X \equiv D^{-1}CA^{-1}B$.

To evaluate this determinant further, let us consider the eigenvalues 
of the Laplacians 
$\hat{\Delta}_0\equiv -2g^{z\zb}\hat{\D}^{(1)}_\zb\hat{\D}^{(0)}_z$ and
$\hat{\Delta}_1\equiv -2g^{z\zb}\hat{\D}^{(0)}_z\hat{\D}^{(1)}_\zb$,
which are acting on 0-form and 1-form eigenfunctions, respectively.
If the 0-form eigenfunctions $g_n$ have non-vanishing eigenvalues, i.e.,
\be
\hat{\Delta}_0 g_n = \Lambda_n g_n,
\nn
\ee
with $\Lambda_n \neq 0$, then there are associated 1-form eigenfunctions,
which also have the same non-vanishing eigenvalues, since
\be
\hat{\Delta}_1 (\hat{\D}^{(0)}_z g_n) = \hat{\D}^{(0)}_z \hat{\Delta}_0 g_n = \Lambda_n (\hat{\D}^{(0)}_z g_n).
\nn
\ee
So we find
\be
{\rm Spec}' \hat{\Delta}_0 = {\rm Spec}' \hat{\Delta}_1,
\nn
\ee
where the {\it prime} denotes that the zero eigenvalues are omitted.
Thus, for the non-zero eigenvalue modes, the 1-loop determinants ${\det D}/{\det A}$ cancel  each other
\be
\frac{\det' D}{\det' A}=
\frac{\sqrt{\prod_n (\Lambda_n + |\phi_0|^2 ) (\Lambda_n + |\phi_0|^2 )}}
{\prod_n (\Lambda_n + |\phi_0|^2 ) }=1,
\nn
\ee
by the bosons and fermions.

On the other hand, for zero eigenvalue modes,
there is no one-to-one correspondence between 0-forms and 1-forms.
If we define the number of zero eigenvalue modes of the operator $\D^{(n)}_\mu$ by
\be
n_0 = \dim \ker \D_z^{(0)}, \quad n_1= \dim \ker \D_\zb^{(1)},
\nn
\ee
the zero eigenvalue modes contributes to the 1-loop determinant via
\be
\frac{\det_0 D}{\det_0 A}=
\left(
\frac{(\bar{\phi}_0)^{n_0}(\phi_0)^{n_1}}{|\phi_0|^{2n_0}}
\right)^{\Nf}.
\nn
\ee
So the 1-loop determinant reduces to
\be
\begin{split}
(\text{1-loop det}) &= \frac{1}{\sdet {\cal M}}\\
&=
\left(
\frac{(\bar{\phi}_0)^{n_0}(\phi_0)^{n_1}}{|\phi_0|^{2n_0}}
\right)^{\Nf}
\times  e^{\Tr \log (1-X)}\\
& = \frac{1}{\phi_0^{\Nf(k+\frac{1}{2}\chi_h)}}e^{\Tr \log (1-X)},
\end{split}
\nn
\ee
where we have used the Hirzebruch-Riemann-Roch index theorem
\be
n_0-n_1 = k +\frac{1}{2}\chi_h,
\ee
and the trace ``$\Tr$'' is taken over all modes and species (flavors) of the fields.

Next let us consider the contribution to the 1-loop determinant from $e^{\Tr\log(1-X)}$.
We first expand
\be
\Tr\log(1-X) = -\Tr X + \cdots,
\nn
\ee
then we have
\be
-\Tr X = -2\Nf \tr \frac{1}{( \hat{\Delta}_0 + |\phi_0|^2  )^2}\left(
\phi_0\eta_0\chi_0 + \bar{\phi}_0 g^{z\zb} \lambda_{0,z}\lambda_{0,\zb}
\right),
\nn
\ee
where ``$\tr$'' means the trace over the modes only (the sum over flavors is already taken)
and we have used the fact that the terms proportional to
\be
\tr \frac{\D_\mu^{(n)}}{( \hat{\Delta}_0+ |\phi_0|^2  )^2},
\nn
\ee
in the trace part, vanish.
Using the heat kernel regularization as explained in Appendix \ref{sc:heat_kernel}, we can evaluate
the trace of the operator:
\be
\tr \frac{1}{(  \hat{\Delta}_0 + |\phi_0|^2  )^2} \left(
\phi_0\eta_0\chi_0 + \bar{\phi}_0 g^{z\zb} \lambda_{0,z}\lambda_{0,\zb}
\right)=
\frac{1}{4\pi|\phi_0|^2}\left(
\phi_0\eta_0\chi_0 +  \bar{\phi}_0\sum_{l=1}^h \lambda^{(l)}_{0}\bar{\lambda}^{(l)}_{0}
\right)+\cdots.
\nn
\ee
Thus we obtain
\be
\Tr\log(1-X) \simeq -\frac{\Nf}{2\pi}\left(
\frac{\eta_0\chi_0}{\bar{\phi}_0}+ \frac{\sum_{l=1}^h \lambda^{(l)}_{0}\bar{\lambda}^{(l)}_{0}}{\phi_0}
\right),
\ee
at the 1-loop level.

Now we can explicitly evaluate the vacuum expectation value of 
the operator $e^{\beta{\cal I}_V(g)}$, which gives the volume of 
the moduli space of the vortices.
First of all, we note here that $e^{\beta{\cal I}_V(g)}$ takes a 
value at the fixed point in the Coulomb branch
\be
\left.e^{\beta{\cal I}_V(g)}\right|_{\text{Coulomb branch fixed point}} = e^{
\beta\phi_0\left(\frac{g^2\zeta}{2}\A -2\pi  k\right)
 -\beta \sum_{l=1}^h \lambda^{(l)}\bar{\lambda}^{(l)}},
\label{Coulomb branch value of the operator}
\ee
in terms of the zero modes,
where ${\cal A}$ is the area of the Riemann surface $\Sigma_h$.
We would like to emphasize here that the coupling $g$ on the right-hand side of
eq.~(\ref{Coulomb branch value of the operator})
is the physical coupling of the vortex system whose volume can be evaluated and
 differs from the critical value coupling $g_0=g_c$ in the $Q$-exact action, i.e.,
we can generally assume
\be
\frac{g^2\zeta}{2}\A -2\pi  k \neq 0,
\ee
even in the Coulomb branch localization.
We just need to insert this fixed point value into the path integral since the operator $e^{\beta{\cal I}_V(g)}$
is $Q$-closed.
Putting together the 1-loop correction of the supersymmetric 
Yang-Mills action and the contribution from the operator $e^{\beta{\cal I}_V(g)}$,
we can evaluate the vacuum expectation value of $e^{\beta{\cal I}_V(g)}$ by an integral over the residual zero modes of the vector multiplet
\be
\begin{split}
\left\langle
e^{\beta{\cal I}_V(g)}
\right\rangle^{g_0=g_c}_k
&=\int \frac{d\phi_0}{2\pi i} \frac{d\bar{\phi}_0}{2\pi i} d\eta_0 d\chi_0 \prod_{l=1}^h d\lambda^{(l)}_{0}d\bar{\lambda}^{(l)}_{0}
 \frac{1}{\phi_0^{\Nf(k+\frac{1}{2}\chi_h)}}\\
&\qquad
\times\exp\left\{
\beta\phi_0\left(\frac{g^2\zeta}{2}{\cal A} -2\pi k\right)
-\frac{\Nf}{2\pi\bar{\phi}_0}\eta_0\chi_0
-\left(\beta+\frac{\Nf}{2\pi  \phi_0}\right)\sum_{l=1}^{h} \lambda^{(l)}\bar{\lambda}^{(l)}
\right\},
\end{split}
\label{zero mode integral}
\ee
where we have normalized the integral measure of $\phi_0$ and $\bar{\phi}_0$, dividing by $2\pi i$.
Using the evaluation of the following integral
\be
\int \frac{d\bar{\phi}_0}{2\pi i \bar{\phi}_0} =  \frac{1}{2},
\ee
with a suitable contour
that contains a pole at the origin,
the integral (\ref{zero mode integral}) reduces to a contour integral of $\phi_0$ only
\be
\left\langle
e^{\beta{\cal I}_V(g)}
\right\rangle^{g_0=g_c}_k
=
\frac{\Nf}{4\pi}
\int_C \frac{d\phi_0}{2\pi i} \frac{\left(\beta+\frac{\Nf}{2\pi \phi_0}\right)^h}{\phi_0^{\Nf(k+\frac{1}{2}\chi_h)}}
e^{\beta\phi_0\left(\frac{g^2\zeta}{2}{\cal A} -2\pi k\right)},
\label{abelian contour integral}
\ee
after integrating out all fermionic zero modes and $\bar{\phi}_0$.
The factor $\left(\beta+\frac{\Nf}{2\pi \phi_0}\right)^h$ comes from the integral of the zero modes of $\lambda$
and $\bar{\lambda}$.
This is a physical derivation of the observations in \cite{Moore:1997dj,Gerasimov:2006zt}.

The integrand in (\ref{abelian contour integral}) contains a 
multiple pole at the origin $\phi_0=0$.
If we consider a small shift of the position of the pole by
\be
\phi_0 \to \phi_0 + \e,
\nn
\ee
$\e$ should satisfy
\be
\mathfrak{Re}\,  \e \geq 0,
\nn
\ee
since 
the operator $e^{\beta{\cal I}_V(g)}$, which originally contains a factor 
\be
e^{-\beta\int_{\Sigma_h}d^2 z \sqrt{g}\, (\Phi+\e) HH^\dag},
\nn
\ee
under the shift of $\Phi$, must converge as well as in the Higgs 
branch integral where $\Phi=0$.

On the other hand, in the Coulomb branch integration,
the factor
\be
e^{\beta(\phi_0+\e)\left(\frac{g^2\zeta}{2}{\cal A} -2\pi k\right)}
\nn
\ee
in the integrand shows that 
\be
\mathfrak{Re} \, \phi_0 < 0 \quad \text{or} \quad  \mathfrak{Re} \, \phi_0 > 0
\nn
\ee
is required if $\left(\frac{g^2\zeta}{2}{\cal A} -2\pi k\right)$ 
is positive or negative respectively, when we add an integration 
contour at sufficiently large values of $|\phi_0|$ in 
upper or lower half plane in order to have a closed contour for 
the integral without changing its values. 

\begin{figure}
\begin{minipage}[b]{.5\linewidth}
\centering\includegraphics[width=5cm]{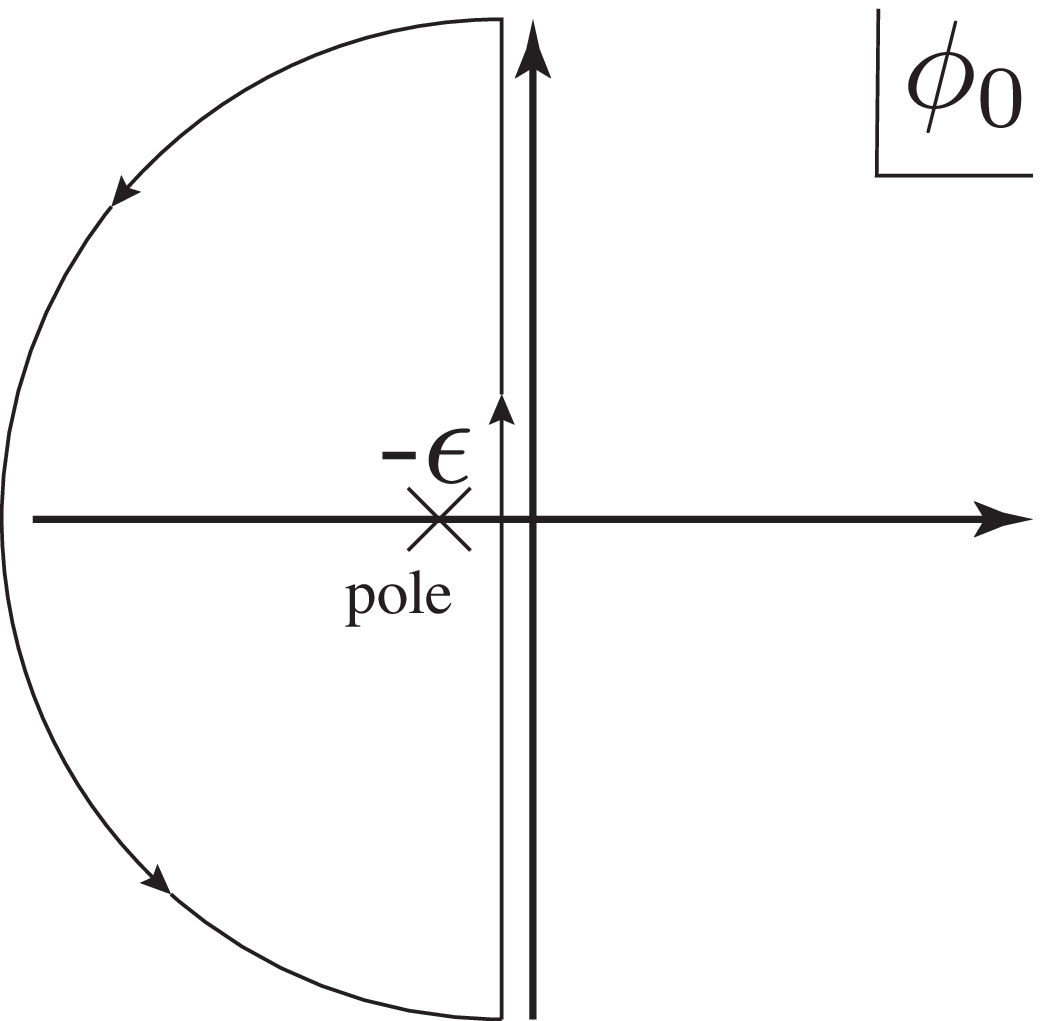}
\subcaption{$\frac{g^2\zeta}{2}{\cal A} -2\pi k >0$}
\end{minipage}%
\begin{minipage}[b]{.5\linewidth}
\centering\includegraphics[width=5cm]{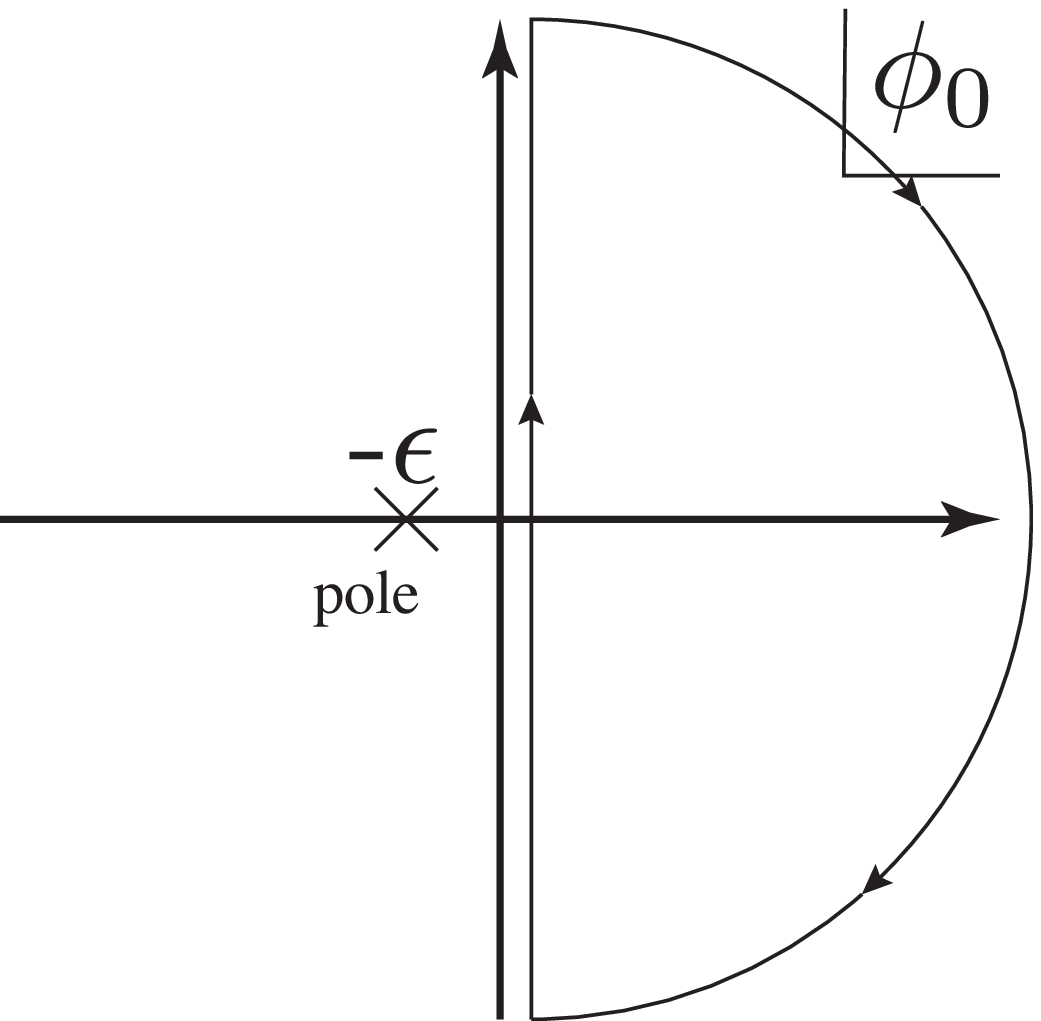}
\subcaption{$\frac{g^2\zeta}{2}{\cal A} -2\pi k<0$}
\end{minipage}
\caption{The choice of contours depending on the sign of 
$\frac{g^2\zeta}{2}{\cal A} -2\pi k$.
The pole is inside the contour
and the volume of the moduli space exists if and only if the Bradlow bound is satisfied.}
\label{contours}
\end{figure}

Thus we can pick up the residues at $\phi_0=-\e$ if and only if
\be
\frac{g^2\zeta}{2}{\cal A} -2\pi k \geq 0.
\label{Bradlow bound}
\ee
(see also Fig.\ref{contours}).
If $\frac{g^2\zeta}{2}{\cal A} -2\pi k$ is negative,
the contour cannot contain the pole and then the integral vanishes. 
This mean that the moduli space (BPS solution) of the vortex 
does not exist if the condition (\ref{Bradlow bound})
is not satisfied. This result is known as the Bradlow bound 
for the vortex \cite{Bradlow:1990ir}. 
The bound can be interpreted as each BPS vortex has an intrinsic 
finite size $4\pi/(g^2\zeta)$ preventing more vortices on the 
Riemann surface of area ${\cal A}$ than ${\cal A}g^2\zeta/(4\pi)$. 
The integral (\ref{zero mode integral}) gives the correct formula 
for the volume of the moduli space without  
explicit knowledge of the moduli space metric. 
It automatically gives the selection rule of integration 
contours leading to the Bradlow bound. 

Let us give a concrete example. If we consider the vortices on the sphere ($h=0$ and $\chi_h=2$),
the contour integral (\ref{abelian contour integral}) gives
\be
\left\langle
e^{\beta{\cal I}_V(g)}
\right\rangle^{g_0=g_c}_k
=\frac{\Nf}{4\pi}\beta^{k\Nf + \Nf -1}\frac{\left(\frac{g^2\zeta}{2}{\cal A} -2\pi k\right)^{k\Nf+\Nf-1}}
{(k\Nf + \Nf -1)!}.
\ee
The power of $\beta$ agrees with the complex dimension of the moduli space
and we find that the volume of the moduli space for the Abelian vortex on the sphere is given by
\be
\Volume {\cal M}^{1,\Nf}_k(S^2;\A) = \frac{\left(\frac{g^2\zeta}{2}{\cal A} -2\pi k\right)^{k\Nf+\Nf-1}}
{(k\Nf + \Nf -1)!},
\ee
up to the irrelevant path integral constant ${\cal N} = \frac{\Nf}{4\pi}$.
For the case of the vortices on the torus ($h=1$),
we obtain
\be
\left\langle
e^{\beta{\cal I}_V(g)}
\right\rangle^{g_0=g_c}_k
=\frac{\Nf}{4\pi}\beta^{k\Nf}\frac{\Nf\frac{g^2\zeta}{4\pi}\A\left(\frac{g^2\zeta}{2}\A -2\pi k\right)^{k\Nf-1}}
{(k\Nf )!},
\ee
and
\be
\Volume {\cal M}^{1,\Nf}_k(T^2;\A) =\frac{\Nf\frac{g^2\zeta}{4\pi}\A\left(\frac{g^2\zeta}{2}\A -2\pi k\right)^{k\Nf-1}}
{(k\Nf )!},
\ee
for $k> 0$.

By setting $\Nf=1$, the above examples agree with \cite{MantonSutcliffe},
where the volume of the moduli space is directly computed from the metrics.

In the case of $k=0$, the contour integral represents
the volume of the vacuum moduli space.
In particular, the contour integral gives
\be
\Volume {\cal M}_0^{1,N_f}(S^2;\A) =  \frac{\left(\frac{g^2\zeta}{2}{\cal A}\right)^{\Nf-1}}
{(\Nf -1)!}=\Volume(\C P^{\Nf-1}),
\ee
which is the volume of the complex projective space with the radius $\frac{g^2\zeta}{4\pi}{\cal A}$,
and
\be
\Volume {\cal M}^{1,\Nf}_0(T^2;\A) =\frac{\Nf}{2\pi},
\ee
which is proportional to the number of isolated vacua (the Witten index).

Finally we comment on the power of $\beta$ (the dimension of the vortex moduli space).
It can be found generally by rescaling of $\phi_0$
as $\phi_0 \to \phi_0'=\phi_0/\beta$ in the integral formula (\ref{abelian contour integral})
\be
\left\langle
e^{\beta{\cal I}_V(g)}
\right\rangle_k
=
\frac{\Nf}{4\pi}
\beta^{\frac{1}{2}(\Nf-1)\chi_h + \Nf k }
\int_C \frac{d\phi'_0}{2\pi i} \frac{\left(1+\frac{\Nf}{2\pi \phi'_0}\right)^h}{{\phi'}_0^{\Nf(k+\frac{1}{2}\chi_h)}}
e^{\phi'_0\left(\frac{g^2\zeta}{2}{\cal A} -2\pi k\right)}.
\nn
\ee
The integral expression here does not depend on $\beta$ any more
and this agrees with the integral formula discussed in \cite{Miyake:2011yr}.
Thus the dimension of the
moduli space of the Abelian vortex is given by
\be
\dim_\C {\cal M}^{1,\Nf}_k(\Sigma_h) = \frac{\chi_h}{2}(\Nf-1) + \Nf k.
\ee
In order for the moduli space to exist, at least the dimension should be equal to or greater than zero\footnote{
If the dimension of the moduli space is zero, the moduli space becomes the zero dimensional
isolated points.
}. So the vorticity is restricted on the generic Riemann surface as
\be
k \geq \max\left(0,-\frac{\chi_h(\Nf-1)}{2\Nf}\right)
=\max\left(0,\frac{(h-1)(\Nf-1)}{\Nf}\right).
\ee
Note here that there is a non-trivial lower bound for the vorticity $k$
on higher genus Riemann surfaces ($\chi_h<0$) if $\Nf>1$,
while the usual bound ($k\geq 0$) holds in the case of
$\Nf=1$ or $\chi_h\geq 0$ ($h=0,1$).
It is interesting to understand this phenomena from the point of view
of the differential equations of the BPS vortex on higher genus Riemann surfaces.


\subsection{Non-Abelian case}

Now we generalize the above localization arguments to the non-Abelian case.

Let us consider the fixed point equation first. The fixed point equations for the non-Abelian theory are given by
\be
\begin{split}
&\D_\mu \Phi = [\Phi,\bar{\Phi}]=0,\\
&\mu_r=\mu_\zb=\mu_z=0.
\end{split}
\ee
Using the Weyl-Cartan bases (see Appendix A), the fixed point equations can be solved by
\be
\begin{split}
&\hat{\Phi} = \phi^i_0\mathtt{H}_i,\\
&\hat{A}_\mu = a^i_\mu \mathtt{H}_i,
\end{split}
\nn
\ee
in a suitable gauge,
where $\phi_0^i$ is a constant zero mode
and the field strength $F^{(i)}_{z\zb}=\del_z a^i_\zb - \del_\zb a^i_z$
gives magnetic fluxes for each $U(1)$ Cartan part
\be
\frac{i}{2\pi}\int_{\Sigma_h} d^2 z \sqrt{g} g^{z\zb} F^{(i)}_{z\zb} = k_i\quad (i=1,\ldots,\Nc),
\nn
\ee
which satisfies $k = \sum_{i=1}^\Nc k_i$.

We now expand fields around the solution of the fixed point equations, i.e.,
\be
\begin{split}
\Phi &= \hat{\Phi} + \frac{1}{\sqrt{t}}\tilde{\Phi}\\
& = \phi^i_0\mathtt{H}_i
+ \frac{1}{\sqrt{t}}\left(\tilde{\phi}^i\mathtt{H}_i + \tilde{\phi}^\alpha \mathtt{E}_\alpha\right),\\
A_\mu & =  \hat{A}_\mu + \frac{1}{\sqrt{t}}\tilde{A}_\mu\\
&= a_\mu^i\mathtt{H}_i
+ \frac{1}{\sqrt{t}}\left(\tilde{A}_\mu^i\mathtt{H}_i + \tilde{A}_\mu^\alpha \mathtt{E}_\alpha\right).
\end{split}
\label{expansion}
\ee
Similarly, the fermions are expanded around the corresponding zero modes:
\be
\begin{split}
\lambda_\mu & = \sum_{l=1}^h \lambda_{0,\mu}^{i,(l)}\mathtt{H}_i+ \frac{1}{\sqrt{t}}\left(
\tilde{\lambda}^i_\mu\mathtt{H}_i + \tilde{\lambda}_\mu^\alpha \mathtt{E}_\alpha
\right)\\
\eta &= \eta_0^i \mathtt{H}_i + \frac{1}{\sqrt{t}}\left(
\tilde{\eta}^i \mathtt{H}_i + \tilde{\eta}^\alpha \mathtt{E}_\alpha
\right),\\
\chi &= \chi_0^i \mathtt{H}_i + \frac{1}{\sqrt{t}}\left(
\tilde{\chi}^i \mathtt{H}_i + \tilde{\chi}^\alpha \mathtt{E}_\alpha
\right).\\
\end{split}
\nn
\ee
Other fields (auxiliary fields $Y$ and the chiral multiplets) are 
expanded around zero, that means just rescaling by
$1/\sqrt{t}$. We omit the {\it tilde} of the fluctuations for 
these fields.

Substituting the above expansion (\ref{expansion}) into the 
Lagrangian, which is also rescaled by ${\cal L}\to t\cal{ L}$, 
we find thanks to the overall coupling independence 
\be
\begin{split}
t{\cal L}_B &= \frac{1}{g_c^2}
\Bigg[
g^{z\zb}\left(
|\del_z \tilde{\phi}^i|^2
+|\del_\zb \tilde{\phi}^i|^2\right)
+\left\{ g^{z\zb}\left(\del_z \tilde{A}^i_\zb - \del_\zb \tilde{A}^i_z\right)\right\}^2\\
&\qquad\quad
+g^{z\zb}\left(|\hat{\D}_z \tilde{\phi}^\alpha-i\alpha(\phi_0)\tilde{A}_z^\alpha|^2
+|\hat{\D}_\zb \tilde{\phi}^\alpha-i\alpha(\phi_0)\tilde{A}_\zb^\alpha|^2\right)\\
&\qquad\quad
+\left\{ g^{z\zb}\left(\hat{\D}_z \tilde{A}^\alpha_\zb - \hat{\D}_\zb \tilde{A}^\alpha_z\right)\right\}^2
+\frac{1}{4}|\alpha(\phi_0)\bar{\tilde{\phi}}^{-\alpha}-\alpha(\bar{\phi}_0)\tilde{\phi}^\alpha|^2
\Bigg]\\
&\qquad\quad
+{\cal O}(1/\sqrt{t})
\end{split}
\nn
\ee
for the bosonic part after eliminating the auxiliary field $Y$, and
\be
\begin{split}
t{\cal L}_F &= \frac{1}{g_c^2}
\Bigg[
2g^{z\zb} \left(
\lambda^i_z \del_\zb \eta^i
-\lambda^i_\zb \del_z \eta^i
+\lambda^{-\alpha}_z \hat{\D}_\zb \eta^\alpha
-\lambda^{-\alpha}_\zb \hat{\D}_z \eta^\alpha
\right)\\
&\qquad\quad
+2ig^{z\zb}\alpha(\bar{\phi}_0)\lambda_z^{-\alpha}\lambda_\zb^\alpha
+i\alpha(\phi_0)\eta^{-\alpha}\eta^\alpha
-i\alpha(\phi_0)\chi^{-\alpha}\chi^\alpha\\
&\qquad\quad
-2g^{z\zb}\left(
\chi^i (\del_z \lambda_\zb^i
+\del_\zb \lambda_z^i)
+\chi^{-\alpha}( \hat{\D}_z\lambda^\alpha_\zb
+\hat{\D}_\zb \lambda^\alpha_z)
\right)
\Bigg]\\
&\qquad\quad
+{\cal O}(1/\sqrt{t})
\end{split}
\nn
\ee
for the fermionic part,
where $\alpha(\phi_0)\equiv \sum_i \alpha_i \phi_0^i$,
$\hat{\D}_z \tilde{\phi}^\alpha \equiv \del_z \tilde{\phi}^\alpha+i\alpha(a_z)\tilde{\phi}^\alpha$, etc.

Introducing two component fermions by
\be
\begin{array}{lcl}
\Psi^i = (\chi^i-\eta^i,-\lambda^i_\zb)^T, &&
\bar{\Psi}^i = (\chi^i+\eta^i,\lambda^i_z),\\
\Psi^\alpha  = (\chi^\alpha+\eta^\alpha,-\lambda^\alpha_\zb)^T, &&
\bar{\Psi}^{-\alpha} = (\chi^{-\alpha}-\eta^{-\alpha},\lambda^{-\alpha}_z),
\end{array}
\nn
\ee
the quadratic Lagrangian of the fermionic part is written simply by
\be
t{\cal L}_F = \frac{1}{g_c^2}
\left[
\sum_i\bar{\Psi}^i \slashed{\del} \Psi^i
+\sum_\alpha\bar{\Psi}^{-\alpha} ( \slashed{\D} - iM_\alpha(\phi_0)) 
\Psi^\alpha
\right]+{\cal O}(1/\sqrt{t}),
\nn
\ee
where
\be
\slashed{\del} = \begin{pmatrix}
0 & 2g^{z\zb}\del_z \\
-2g^{z\zb}\del_\zb & 0
\end{pmatrix},
\quad
\slashed{\D} = \begin{pmatrix}
0 & 2g^{z\zb}\hat{\D}_z \\
-2g^{z\zb}\hat{\D}_\zb & 0
\end{pmatrix},
\quad
M_\alpha(\phi_0) = \begin{pmatrix}
\alpha(\phi_0) & 0\\
0 & 2g^{z\zb}\alpha(\bar{\phi}_0)
\end{pmatrix}.
\nn
\ee
At a generic value of $\phi_0$, the root components of the fermions 
such as $\Psi^\alpha$ are always massive.
Thus there is no true zero mode in the off-diagonal components 
and we expect that there are zero modes only in the Cartan part 
of the fermions.

Because of the supersymmetry, we can expect essentially that 
the 1-loop determinants reduce to one by cancellation of
bosons and fermions for the non-zero modes.
We should, however, pay attention to zero eigenvalue states 
of the operator $\hat{\D}_\mu$.
According to the index theorem on the Riemann surface $\Sigma_h$, 
the numbers of the zero eigenvalue state for 0-forms and 1-forms 
on $\Sigma_h$ differ. So the contributions to 1-loop determinant 
from these zero eigenvalue states should not cancel each other.
We call these zero eigenvalue states pseudo-zero modes.

Actually, if we evaluate the 1-loop determinant from
the off-diagonal components of the pseudo-zero modes, it reduces to
\be
\prod_{i \neq j}(\phi_0^i-\phi_0^j)^{\frac{1}{2}\chi_h+k_i-k_j}=(-1)^\sigma
\prod_{i < j}(\phi_0^i-\phi_0^j)^{\chi_h},
\ee
where
we have used the Hirzebruch-Riemann-Roch index theorem for $\slashed{\D}$ and
 $(-1)^\sigma$ is a sign factor depending on the total magnetic flux $k$ via
\be
\begin{split}
(-1)^\sigma &= (-1)^{\frac{\chi_h}{4}\Nc(\Nc-1) - \sum_{i<j}(k_i-k_j)}\\
&=
\begin{cases}
(-1)^{\frac{\chi_h}{4}\Nc(\Nc-1)+k} & \text{if $\Nc$ is even}\\
(-1)^{\frac{\chi_h}{4}\Nc(\Nc-1)} & \text{if $\Nc$ is odd}
\end{cases}.
\end{split}
\ee

After integrating out all off-diagonal components of the fields,
the argument of the localization for each Cartan part is almost parallel to the Abelian case in the previous subsection.
Using the $Q$-exactness of each Abelian component of the effective $U(1)^\Nc$ theory, we can vary the
$i$-th $U(1)$ gauge coupling $g_0^{(i)}$ to be independently $g^{(i)}_0=g^{(i)}_c$ satisfying
\be
(g^{(i)}_c)^2\zeta = \frac{4\pi k_i}{\A}
\ee

Choosing the Lorentz gauge
$\del^\mu a^i_\mu = 0$
for each $U(1)$ part, the vacuum expectation value of the operator $e^{\beta{\cal I}_V(g)}$
with a fixed partition of $k$ reduces to the zero mode integral
\be
\begin{split}
\left\langle
e^{\beta{\cal I}_V(g)}
\right\rangle^{\vec{g}_0=\vec{g}_c}_{\vec{k}}
&=(-1)^\sigma\int  \prod_{i=1}^{\Nc}\frac{d\phi^i_0}{2\pi i} \frac{d\bar{\phi}^i_0}{2\pi i}
 d\eta^i_0 d\chi^i_0 \prod_{l=1}^h d\lambda^{i,(l)}_{0}d\bar{\lambda}^{i,(l)}_{0}
 \frac{1}{(\phi^i_0)^{\Nf(k_i+\frac{1}{2}\chi_h)}}\prod_{i<j}(\phi_0^i-\phi_0^j)^{\chi_h}\\
&\qquad
\times\exp\left[
\sum_{i=1}^{\Nc}\left\{
\beta\phi^i_0\left(\frac{g^2\zeta}{2}{\cal A} -2\pi k_i\right)
-\frac{\Nf}{2\pi i \bar{\phi}^i_0} \eta^i_0\chi^i_0
-\left(\beta+\frac{\Nf}{2\pi \phi^i_0}\right)\sum_{l=1}^{h} \lambda^{i,(l)}_0 \bar{\lambda}^{i,(l)}_0
\right\}
\right],
\end{split}
\nn
\ee
where $\vec{g}_0=\vec{g}_c$ stands for $g^{(i)}_0=g^{(i)}_c$ $(i=1,\ldots,\Nc)$ and
$\vec{k}=(k_1,\ldots,k_\Nc)$.
Here we again note that the coupling $g$ on the right-hand side coming from the inserted operator
differs from the coupling $g^{(i)}_0=g^{(i)}_c$ in the $Q$-exact action.
After integrating out the fermionic zero modes and $\bar{\phi}^i_0$'s, 
the path integral finally reduces to
a contour integral formula
\be
\begin{split}
\left\langle
e^{\beta{\cal I}_V(g)}
\right\rangle^{\vec{g}_0=\vec{g}_c}_{\vec{k}}
&=
\left(\frac{\Nf}{4\pi}\right)^{\Nc} 
(-1)^\sigma\int_C \prod_{i=1}^{\Nc} \frac{d\phi^i_0}{2\pi}\prod_{i<j}(\phi_0^i-\phi_0^j)^{\chi_h}\\
&\qquad\qquad\qquad\qquad\qquad\qquad
\times \prod_{i=1}^{\Nc}
\frac{\left(\beta+\frac{\Nf}{2\pi \phi^i_0}\right)^h}{(\phi^i_0)^{\Nf(k_i+\frac{1}{2}\chi_h)}} e^{\beta\sum_{i=1}^{\Nc}\phi^i_0\left(\frac{g^2\zeta}{2}{\cal A} -2\pi k_i\right)}.
\end{split}
\nn
\ee
By summing over the partition of the total vorticity $k=\sum_{i=1}^{\Nc}k_i$ into $\vec{k}=(k_1,k_2,\ldots,k_\Nc)$, we obtain 
the integral formula for the volume of the moduli space of the 
non-Abelian vortices
\begin{multline}
\beta^{\dim_\C {\cal M}_k}\Volume {\cal M}_k\\
=
(-1)^\sigma \sum_{|\vec{k}|=k}\int_C \prod_{i=1}^{\Nc} \frac{d\phi^i_0}{2\pi i}\prod_{i<j}(\phi_0^i-\phi_0^j)^{\chi_h}
\prod_{i=1}^{\Nc}
\frac{\left(\beta+\frac{\Nf}{2\pi \phi^i_0}\right)^h}{(\phi^i_0)^{\Nf(k_i+\frac{1}{2}\chi_h)}}
e^{\beta\sum_{i=1}^{\Nc}\phi^i_0\left(\frac{g^2\zeta}{2}{\cal A} -2\pi k_i\right)},
\label{non-Abelian contour integral}
\end{multline}
up to the irrelevant numerical constant ${\cal N}\equiv \left(\frac{\Nf}{4\pi}\right)^{\Nc}$.
This is the contour integral expression of the volume of the non-Abelian vortex moduli space and agrees with our previous result \cite{Miyake:2011yr} by setting $\beta=1$.

The power of $\beta$ is also easily found by rescaling $\phi^i_0$ as
$\phi^i_0 \to {\phi'}^i_0=\phi^i_0/\beta$.
Using this rescaling, we find that the dimension of the moduli space of the non-Abelian vortex
is generally given by
\be
\begin{split}
\dim_\C {\cal M}^{\Nc,\Nf}_k(\Sigma_h)
&= \frac{\chi_h}{2} \Nc(\Nf-\Nc) + \Nf k\\
&= \frac{\chi_h}{2} \Nc\tilde{N}_C + (\Nc+{\tilde{N}}_C)k,
\end{split}
\label{dimension of non-Abelian vortex}
\ee
where $\tilde{N}_C\equiv \Nf-\Nc$.
It is interesting that the dimension
of the moduli space (\ref{dimension of non-Abelian vortex}) is
invariant under the duality transformation $\Nc \leftrightarrow \tilde{N}_C$.
The $k$-dependent part in (\ref{dimension of non-Abelian vortex})
agrees with the result on the flat space $\R^2$ \cite{Hanany:2003hp,Eto:2005yh,Eto:2006pg}.
The positivity of the dimension leads to the lower bound of the vorticity
\be
k\geq 
\max\left(0,
(h-1)\frac{\Nc\tilde{N}_C}{\Nc+\tilde{N}_C}
\right).
\label{non-Abelian lower bound}
\ee
From the viewpoint of the BPS differential equations,
it is difficult 
to find a topology ($\chi_h$) dependent part
in (\ref{dimension of non-Abelian vortex})
or (\ref{non-Abelian lower bound}),
that also satisfies the duality.


\subsection{Bradlow bound and Jeffrey-Kirwan residue formula}

The contour integral (\ref{abelian contour integral}) by $\phi_0$ has non-vanishing residue if and only if
\be
\frac{g^2\zeta}{2}{\cal A} -2\pi k \geq 0
.
\label{non-vanishing residue condition}
\ee
The 
condition is known as the Bradlow bound which immediately follows 
from 
the BPS equations
\be
\begin{split}
\int_{\Sigma_h}d^2 z\sqrt{g} \left(
\frac{g^2\zeta}{2}-ig^{z\zb}{F_{z\zb}}
\right)
&=\frac{g^2\zeta}{2}\A - 2 \pi k\\
&= \frac{g^2}{2}\int_{\Sigma_h}d^2 z\sqrt{g}\langle HH^\dag\rangle\geq 0.
\end{split}
\nn
\ee

A similar selection rule for the contour is also known as the Jeffrey-Kirwan residue formula
\cite{Jeffrey-Kirwan},
in mathematical literature, to satisfy the D-term condition
\be
\langle HH^\dag \rangle = \zeta.
\nn
\ee
The contour for the Jeffrey-Kirwan residue formula is chosen
to get  non-vanishing and vanishing residues if and only if $\zeta\geq 0$ and $\zeta < 0$, respectively.
This Jeffrey-Kirwan residue formula causes  wall-crossing phenomena in supersymmetric quantum mechanics.

The Bradlow bound can be considered as a generalization of 
the Jeffrey-Kirwan residue formula for the effective FI parameter
including the magnetic flux
\be
\langle HH^\dag \rangle = \zeta_{\rm eff}(k),
\nn
\ee
where
\be
\zeta_{\rm eff}(k) = \zeta - \frac{4\pi}{g^2}\frac{k}{\A},
\ee
is a function of the number density of the vortex $\rho = k/\A$.
The contour is chosen whether  $\zeta_{\rm eff}(k)$ is positive or not.

For the non-Abelian theory,
our integral formula  (\ref{non-Abelian contour integral}) suggests
the effective FI parameter for each Abelian part as
\be
\zeta_{\rm eff}(k_i)  = \zeta - \frac{4\pi}{g^2}\frac{k_i}{\A}.
\ee
This is also a generalization of the Jeffrey-Kirwan residue formula in the non-Abelian gauge theories.


\section{Generating Function}
\label{sc:generating_func}

So far, we have considered the volume of the moduli space under 
a fixed magnetic flux $k$.
We consider the generating function of the volume of the moduli 
space, which can be obtained by a summation over the flux $k$
\be
Z_{\Nc,\Nf}(q;\Sigma_h) = \sum_k \beta^{\dim_\C {\cal M}_k}\Volume {\cal M}_k \, q^k,
\label{generating function}
\ee
where we set $q = e^{-2\pi \tau}$.
$Z_{\Nc,\Nf}$ can be regarded as the field theoretical partition 
function (\ref{field theory partition function}) with the insertion 
of the operator $e^{\beta {\cal I}_V(g)}$.
We  should, however, note that the summation over $k$ in the 
generating function (\ref{generating function}) is restricted 
from above by the Bradlow bound, which depends on the size ${\cal A}$ 
of the Riemann surface.

Under this restriction of the summation over $k$, the explicit 
evaluation of the generating function (\ref{generating function}) 
is rather difficult. 
So we consider only the case that the area of the Riemann surface 
$\A$ or the physical couplings $g^2 \zeta$ are sufficiently large, namely,
cases where we can take the summation up to $k\to \infty$. 
This implies that we should use the integration contour to enclose the pole at $\phi_0=-\epsilon$ 
as in Fig.~\ref{contours} (a). 
We note that we can shift the position of the pole to the left at finite 
distances away from the origin without modifying the result. 

\begin{figure}
\centering\includegraphics[width=5cm]{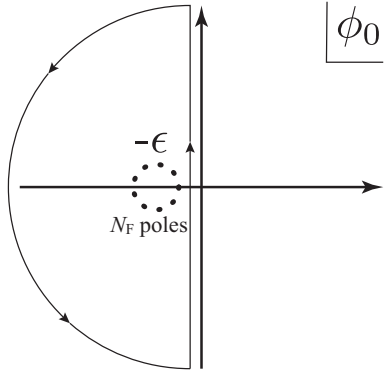}
\caption{The integral contour of the generating function and a split of the degenerated pole.
After summing up the vorticity $k$ ignoring the Bradlow bound, the degenerated pole splits into $\Nf$ simple poles.
The integral contour still encloses all the poles
since we can shift the center of the poles by sufficiently large distance $\epsilon$
against the distributions of the poles.
}
\label{generating function contour}
\end{figure}

Let us see some concrete examples.
For the Abelian theory ($G=U(1)$), the generating function is given by
\be
Z_{1,\Nf}(q;\Sigma_h) = \sum_{k=0}^{\infty} \int_C \frac{d\phi_0}{2\pi i}
\frac{\left(\beta+\frac{\Nf}{2\pi \phi_0}\right)^h}
{\phi_0^{\Nf(k+\frac{1}{2}\chi_h)}}
e^{2\pi \beta\phi_0\left(\hat{\A}- k\right)}q^k,
\ee
where the contour $C$ is always chosen to enclose the pole at 
$\phi_0=-\epsilon$, and 
\be
\hat{\A}\equiv \frac{g^2\zeta}{4\pi}\A.
\nn
\ee
If we take the summation of $k$ first assuming the interchangeability 
of sum and integral, we obtain 
\be
\begin{split}
Z_{1,\Nf}(q;\Sigma_h) &= \int_C \frac{d\phi_0}{2\pi i} \frac{\left(\beta+\frac{\Nf}{2\pi \phi_0}\right)^h}{\phi_0^{\frac{\chi_h}{2}\Nf}}
e^{2\pi\beta\phi_0\hat{\A}} \sum_{k=0}^{\infty} \left( \frac{q}{\phi_0^\Nf e^{2\pi\beta\phi_0}} \right)^k\\
&= \int_{C'} \frac{d\phi_0}{2\pi i} \frac{\left(\beta \phi_0^\Nf+\frac{\Nf}{2\pi}\phi_0^{\Nf-1}\right)^h}{\phi_0^\Nf-q e^{-2\pi \beta \phi_0}}
e^{2\pi\beta\phi_0\hat{\A}}.
\end{split}
\label{GF contour integral}
\ee
The integrand of the above contour integral has poles at zeros 
of the denominator, which are solutions of
\be
\phi_0^\Nf-q e^{-2\pi \beta \phi_0} = 0.
\label{zeros equation}
\ee
The original degenerated pole at $\phi_0=-\epsilon$ spreads out into $\Nf$ simple poles, which
are distributed roughly in the range of $|q|^{1/\Nf}$.
The integration contour $C'$ still encloses all the above poles 
since we can shift the center of the poles by a (sufficiently 
large) finite distance $\epsilon$ away from the origin.
(see Fig.~\ref{generating function contour}).

There is no analytical solution of the transcendental equation 
(\ref{zeros equation}), but we have generally $\Nf$ independent 
solutions denoted by $x^*_a$ ($a=1,\ldots,\Nf$).
In terms of these solutions, the contour integral 
(\ref{GF contour integral}) can be rewritten as 
\be
\begin{split}
Z_{1,\Nf}(q;\Sigma_h) 
&= \int_C \frac{d\phi_0}{2\pi i} \frac{\left(\beta \phi_0^\Nf+\frac{\Nf}{2\pi}\phi_0^{\Nf-1}\right)^h}
{\prod_{a=1}^\Nf(\phi_0-x_a^*)}
e^{2\pi\beta\phi_0\hat{\A}}\\
&= \sum_{a=1}^\Nf 
\frac{\left(\beta +\frac{\Nf}{2\pi x_a^*}\right)^h}
{\prod_{b\neq a}(x_a^*-x_b^*)}
e^{2\pi\beta x_a^* (\hat{\A}-h)}q^h.
\end{split}
\label{generic residue sum}
\ee

If we assume $x_a^*\ll1$, which corresponds to $q\ll1$ ($\tau\to\infty$),
then $x^*_a$ approximately becomes
\be
x^*_a \simeq \omega^{a-1} q^{1/\Nf},
\label{approximation of x_a}
\ee
where $\omega=\exp\left( \frac{2\pi i}{\Nf}\right)$ is $\Nf$-th root of unity.
Plugging this approximation into (\ref{generic residue sum}), we obtain
\be
\begin{split}
Z_{1,\Nf}(q;\Sigma_h) 
&\simeq \sum_{a=1}^\Nf 
\frac{\left(\beta +\frac{\Nf}{2\pi \omega^{a-1} q^{1/\Nf}}\right)^h}
{\prod_{b\neq a}(\omega^{a-1}-\omega^{b-1})}
e^{2\pi\beta \omega^{a-1} q^{1/\Nf} (\hat{\A}-h)}q^{h-\frac{\Nf-1}{\Nf}}\\
&=
\sum_{j=0}^h \binom{h}{j}\beta^{j}\left(\frac{\Nf}{2\pi} \right)^{h-j}
\sum_{l=1}^\infty\frac{1}{l!}\left( 2\pi\beta (\hat{\A}-h) \right)^l\\
&\qquad\qquad \times
\frac{1}{\Nf}\sum_{a=1}^{\Nf}\omega^{(j+l+1-h)(a-1)}
q^{h-1+\frac{j+l+1-h}{\Nf}},
\end{split}
\label{u1 q series}
\ee
where we have used the identity
\be
\prod_{b\neq a}(\omega^{a-1}-\omega^{b-1})
=\lim_{x\to \omega^{a-1}}\left(\frac{x^\Nf-1}{x-\omega^{a-1}}\right)=\frac{\Nf}{\omega^{a-1}}.
\nn
\ee
Because of the identity 
\be
\frac{1}{\Nf}\sum_{a=1}^{\Nf}\omega^{(j+l+1-h)(a-1)}=
\begin{cases}
1 & \text{if} \quad j+l+1-h \equiv 0 \pmod{\Nf}\\
0 & \text{otherwise}
\end{cases},
\nn
\ee
we find that the power of $q$ in (\ref{u1 q series})
\be
k = h - 1+\frac{j+l+1-h}{\Nf}
\nn
\ee
always becomes an integer number with a bound
\be
d(k) \equiv k\Nf-(h-1)(\Nf-1) = j+l \geq 0,
\ee
i.e., $k\geq \max\left(-\frac{(1-h)(\Nf-1)}{\Nf},0\right)$.
This condition is nothing but the positivity of the dimension of the moduli space.
Furthermore, using a bound
\be
d(k)-j = l \geq 0,
\nn
\ee
we find
\be
0 \leq j \leq \min\left(h,d(k)\right) 
\nn
\ee

Using these definitions and bounds, we can rewrite Eq.~(\ref{u1 q series}) as
\be
\begin{split}
Z_{1,\Nf}(q;\Sigma_h) 
&=\sum_{k=k_0}^\infty \beta^{d(k)}
\sum_{j=0}^{\min(h,d(k))} \binom{h}{j}\left(\frac{\Nf}{2\pi} \right)^{h-j}
\frac{\left(2\pi(\hat{\A}-h) \right)^{d(k)-j}}{(d(k)-j)!}
q^{k},
\end{split}
\nn
\ee
where
\be
k_0\equiv \left\lceil\frac{(h-1)(\Nf-1)}{\Nf}\right\rceil,
\label{lower bound of vorticity}
\ee
using the ceiling function.
Thus we find the volume of the moduli space
\be
\beta^{\dim_\C {\cal M}_k}\Volume{\cal M}_k \simeq \beta^{d(k)}
\sum_{j=0}^h \binom{h}{j}\left(\frac{\Nf}{2\pi} \right)^{h-j}
\frac{\left(2\pi\hat{\A} \right)^{d(k)-j}}{(d(k)-j)!},
\nn
\ee
in the large area limit $\hat{\A}/k\to \infty$ for fixed $k$.
This agrees with our previous result \cite{Miyake:2011yr},
and the power of $\beta$, $d(k)$ represents the dimension of the moduli space as we expected in the 
Higgs branch analysis.

Next let us consider the non-Abelian case.
Ignoring the Bradlow bound, we take the summation over the vorticity 
first, then we have
\be
\begin{split}
Z_{\Nc,\Nf}(q;\Sigma_h)
&=(-1)^{\frac{\chi_h}{4}\Nc(\Nc-1)}\int_C \prod_{i=1}^{\Nc} \frac{d\phi^i_0}{2\pi i}\prod_{i<j}(\phi_0^i-\phi_0^j)^{\chi_h}\\
&\qquad\qquad\qquad \times
\prod_{i=1}^{\Nc}
\frac{\left(\beta (\phi_0^i)^\Nf+\frac{\Nf}{2\pi}(\phi_0^i)^{\Nf-1}\right)^h}
{(\phi^i_0)^\Nf\mp q e^{-2\pi\beta \phi_0^i}}
e^{2\pi\beta\hat\A \phi^i_0},
\end{split}
\label{non-Abelian generating function}
\ee
where the sign in front of $q$ depends on whether $\Nc$ is even or odd.

Again, if we denote a set of solutions to the transcendental equation
\be
(\phi^i_0)^\Nf\mp q e^{-2\pi\beta \phi_0^i}=0,
\label{transcendental equation}
\ee
by $x^*_a$ ($a=1,\ldots,\Nf$),
the contour integral (\ref{non-Abelian generating function}) is evaluated in terms of the residues
\be
\begin{split}
Z_{\Nc,\Nf}(q;\Sigma_h)
&=(-1)^{\frac{\chi_h}{4}\Nc(\Nc-1)} \Nc!
\sum_{\{x^*_{a_i}\}}
\prod_{i<j}(x^*_{a_i}-x^*_{a_j})^{\chi_h}\\
&\qquad\qquad\qquad \times
\prod_{i=1}^{\Nc}
\frac{\left(\pm\beta +\frac{\Nf}{2\pi x^*_{a_i}}\right)^h}
{\prod_{a\neq a_i}(x^*_{a_i}-{x^*_a})}
e^{2\pi\beta(\hat\A-h) x^*_{a_i}}q^h,
\end{split}
\label{non-Abelian residue expression}
\ee
where the $a_i$ are a set of $\Nc$ indices chosen from $\Nf$ indices $a$,
and ordered as
$a_1 < a_2 < \cdots < a_{\Nc}$. (Note that we are assuming $\Nc\leq \Nf$.)
This choice of indices comes from the fact that 
we can rearrange the order of the indices up to the Weyl 
permutation of the gauge group, whereas the Vandermonde 
determinant necessitates the choice of different poles 
for different $\phi_0^i$ integrals. 
Thus we have summation over the set of indices, whose number 
is given by $\binom{\Nf}{\Nc}$ in total.

It is difficult to evaluate further the expression of the volume (\ref{non-Abelian residue expression}),
since the transcendental equation (\ref{transcendental equation}) does not have analytic solutions
in general, but the case of $\Nc=\Nf=N$ and $h=0$ ($\Sigma_h=S^2$), i.e.,
the non-Abelian local vortex on the sphere, is rather special.
Indeed, in this case, the Vandermonde determinant is divisible by the denominator in
(\ref{non-Abelian residue expression}), and it reduces to
\be
Z_{N,N}(q;S^2)
=N! \, 
e^{2\pi\beta \hat\A \sum_{a=1}^N x^*_{a}},
\label{non-Abelian local vortex case}
\ee
where the sign factor also disappears by a cancellation with the divisor.

If we use the approximation (\ref{approximation of x_a}), we cannot obtain the $q$-dependence of the
generating function since $\sum_{a=1}^N \omega^{a-1}=0$. So we need the approximation to the next order by
\be
x_a^* \simeq \omega^{a-1}(\pm q)^{1/N}e^{-2\pi/N \beta \omega^{a-1}(\pm q)^{1/N}}.
\nn
\ee
Using this approximation, we find
\be
\begin{split}
\sum_{a=1}^N x^*_a &\simeq \sum_{a=1}^N \omega^{a-1}(\pm q)^{1/N}
\sum_{l=0}^\infty \frac{1}{l!} \left(-\frac{2\pi\beta}{N} \right)^l \omega^{l(a-1)}(\pm q)^{l/N}\\
& = \frac{(2\pi\beta)^{N-1}}{(N-1)!}q+{\cal O}(q^2).
\end{split}
\nn
\ee
Substituting this approximation into (\ref{non-Abelian local vortex case}),
the generating function of the volume of the non-Abelian local vortex is given by
\be
Z_{N,N}(q;S^2)
\simeq N! \times 
\sum_{k=0}^\infty \frac{1}{k!}\left\{ \frac{(2\pi\beta)^N}{(N-1)!} \hat{\A} \right\}^k q^k.
\ee
So we find that the volume of the moduli space of the non-Abelian 
local vortex
becomes
\be
\beta^{\dim_\C {\cal M}_k}
\Volume {\cal M}^{N,N}_k(S^2) \simeq \frac{N!}{k!}\left\{ \frac{(2\pi\beta)^N}{(N-1)!} \hat{\A} \right\}^k,
\label{non-Abelian local vortex volume}
\ee
in the large area limit.

This volume of the moduli space of the non-Abelian local vortex 
has been conjectured in eq.~(4.52) of \cite{Miyake:2011yr}
by inference from the concrete evaluation for the $N=2,3$ cases, 
but the conjecture turns out to be in slight disagreement with 
our present result (\ref{non-Abelian local vortex volume}), which 
shows a slightly different coefficient.
We have derived, for the general $N$, the reduction of (the dimension of) 
the volume of the local vortex moduli space, where the moduli space 
volume is proportional to $\hat{\A}^k$ rather than $\hat{\A}^{kN}$, 
by using the generating function.
This is one of the advantages of using the generating function of 
the volume of the vortex moduli space.

\section{Conclusion and Discussion}
\label{sc:discussion}

In this paper, we derive an integral formula for the volume of the moduli space of the BPS vortex on the closed Riemann surface
with the arbitrary genus. The BPS vortex system is embedded 
into ${\cal N}{=}(2,2)$ supersymmetric Yang-Mills theory with matters,
where we have used natural topological twisting on the curved space
by turning on the background flux of the gauged $R$-symmetry.
The background flux is compatible with the BPS vortex and
preserves just half of the supercharges
while the other half of the supercharges is preserved by
the background for the anti-BPS vortex.
This means that the zero BPS vortex sector (vacuum)
on the Riemann surface differs
from the zero anti-BPS vortex sector except on the torus ($h=1$).

We firstly find that the path integral of the supersymmetric 
Yang-Mills theory in the Higgs branch
gives directly the integral over the vortex moduli space. So the partition function of the supersymmetric Yang-Mills theory 
essentially gives the volume of the moduli space except for the integration of fermionic zero modes.
Due to the fermionic zero modes, the partition function itself vanishes.
We need to insert the appropriate operator in order to obtain the moduli space volume from the path integral.
The inserted operator just compensates the fermionic zero modes
and reduces to unity at the localization fixed point.

Secondly, in the Higgs branch description, we cannot perform the moduli space integral
since the metric of the moduli space is not known
in general. However, if we evaluate the same supersymmetric system in the Coulomb branch description
by using the localization method,
the path integral reduces to a simple finite-dimensional contour integral,
which should give the volume of the vortex moduli space
as discussed in the Higgs branch description. 
We also derive the exact 1-loop contribution to the gaugino mass including the higher genus case,
which is needed to make the effective action supersymmetric.

The localization formula for the vortex moduli space
captures the effect of the finite area of the Riemann surface, known as
the Bradlow bound.
The choice of the contours changes whether the area and vorticity satisfy the bound or not.
This can be regarded as a kind of wall-crossing or Jeffrey-Kirwan 
residue formula
where the choice of the contour depends on the flux in general. 

We also discussed the generating function of the volume of the moduli space of the vortex.
Under some assumptions, we can take the summation over the vorticity first.
The summation modifies the contour integral whose poles and residues are 
given 
by the transcendental equations and are difficult to obtain analytically.
However, 
this generating function can give a simple understanding
of the reduction of the moduli space dimension
in the case of the local vortices ($\Nc=\Nf=N$).

Our volume formula for the vortex moduli space on the Riemann surface
suggests that there is a lower bound of vorticity
(\ref{lower bound of vorticity}) on a Riemann surface with
a higher genus ($h>1$ and $\Nf>\Nc$),
besides the upper Bradlow bound.
This means that
there is no solution to the BPS vortex equations for too few vortices
on a higher genus surface.
It is interesting to understand this from the point of view of the
BPS differential equations by using the moduli matrix method \cite{Eto:2006pg},
or the Jacobian variety of
the Riemann surface \cite{Manton:2010sa,Romao:2012vy}.

In this paper, we consider only the case of the closed Riemann surface.
If there are boundaries (punctures) of the Riemann surface,
we should consider holonomies of the gauge fields around the boundaries.
We expect that the partition function (the volume of the vortex moduli space)
is a function of the boundary holonomies besides the vorticity and area.
As known from \cite{Cordes:1994fc},
the partition function of the pure bosonic Yang-Mills theory on the arbitrary
punctured Riemann surface can be constructed from those on 
one, two and three punctured spheres (disk, cylinder, pants)
by gluing together at some boundaries.
So we can expect that the volume of the vortex moduli space
on the punctured Riemann surfaces may also be constructed 
from similar building blocks.

Our system and evaluations can be extended to three dimensions, 
like $S^1\times \Sigma_h$ \cite{Ohta:2012ev,Benini:2015noa,Closset:2016arn}.
The operator which measures the volume of the vortex moduli space
naturally uplifts to the Chern-Simons operator in three-dimensions.
So if we consider Yang-Mills-Chern-Simons-matter theory
in three-dimensions, the partition function may give a counterpart
of the volume of the moduli space of the vortex.
After summing up the vorticity in the Yang-Mills-Chern-Simons-matter theory,
the Bethe equations appear \cite{Okuda:2012nx,Kanno:2018qbn} to determine
the position of the poles in the contour integral
as a generalization of our transcendental equations.
In these analyses, the effects of the size of the vortices do not appear.
So it is interesting to consider the dependence on the finite area $\A$
of the Riemann surface to these three-dimensional theories.

\section*{Acknowledgements}
We would like to thank 
Toshiaki Fujimori,
So Matsuura,
Keisuke Ohashi,
Yuya Sasai
and Yutaka Yoshida
for useful discussions and comments. 
One of the authors (N.\ S.) thanks Takuya Okuda for useful 
discussion. 
 
This work is supported in part 
by Grant-in-Aid for Scientific Research (KAKENHI) (C) 
 Grant Numbers 26400256 and 17K05422 (K.\ O.),  
by 
Grant-in-Aid for Scientific Research (KAKENHI) (B) Grant Number 
18H01217 (N.\ S.), and by the Ministry of Education, Culture, 
Sports, Science, and Technology(MEXT)-Supported Program for the 
Strategic Research Foundation at Private Universities ``Topological 
Science" (Grant No. S1511006) (N.\ S.).

\appendix

\section{Cartan-Weyl basis}
\label{sc:cartan_weyl}

An $\Nc\times \Nc$ matrix $X$ in the adjoint representation of $U(\Nc)$ can be
expanded by the so-called Cartan-Weyl bases by
\be
X = \sum_{i=1}^{\Nc} X^i \mathtt{H}_i + \sum_{\alpha} X^\alpha \mathtt{E}_\alpha,
\nn
\ee
where $\alpha$ stands for the root.
The Cartan-Weyl bases satisfy the following algebra
\be
\begin{split}
&[\mathtt{H}_i,\mathtt{H}_j]=0,\\
&[\mathtt{H}_i,\mathtt{E}_{\pm \alpha}]=\pm \alpha_i \mathtt{E}_{\pm\alpha},\\
&[\mathtt{E}_\alpha,\mathtt{E}_{-\alpha}]=\sum_{i=1}^N \alpha_i \mathtt{H}_i,
\qquad [\mathtt{E}_\alpha,\mathtt{E}_\beta] = N_{\alpha,\beta}\mathtt{E}_{\alpha+\beta},\\
&\mathtt{E}_\alpha^\dag =\mathtt{E}_{-\alpha},
\qquad \Tr \mathtt{E}_\alpha\mathtt{E}_\beta = \delta_{\alpha+\beta,0},
\qquad \Tr \mathtt{H}_i \mathtt{H}_j  = \sum_\alpha \alpha_i \alpha_j= \delta_{ij}.
\end{split}
\ee
We use these notations in this paper.

\section{Heat Kernel Regularization}
\label{sc:heat_kernel}

To compute the 1-loop contributions to the fermion bi-linears, we need to consider
the contribution from the propagators in the boson-fermion loop
\be
\tr \frac{1}{(  \hat{\Delta}_0+ |\phi_0|^2  )^2}{\cal O},
\ee
where $\hat{\Delta}_0\equiv -2g^{z\zb}\hat{\D}^{(1)}_z\hat{\D}^{(0)}_\zb$
is a Laplacian acting on the 0-form wave function,
and ${\cal O}(z)$ is an operator. 
The trace is evaluated as an integral over the coordinate  $z$
\be
\begin{split}
\tr \frac{1}{(  \hat{\Delta}_0+ |\phi_0|^2  )^2}{\cal O}
&=\int d^2 z \langle z | \frac{1}{(  \hat{\Delta}_0+ |\phi_0|^2  )^2} | z\rangle {\cal O}(z)\\
&= \int d^2 z \int_0^\infty dt \, t  \langle z |  e^{-t(\hat{\Delta}_0+ |\phi_0|^2)}| z\rangle{\cal O}(z).
\end{split}
\nn
\ee

We need to evaluate essentially
\be
\langle z | e^{-t \hat{\Delta}_0} | z \rangle = \lim_{w\to z} \langle z | e^{-t \hat{\Delta}_0} | w \rangle,
\label{trace}
\ee
via the heat kernel
\be
h(z,w;t) = \langle z | e^{-t\hat{\Delta}_0} | w \rangle.
\nn
\ee
The heat kernel $h(z,w;t)$ obeys the heat equation
\be
\left(\frac{\del}{\del t} + \hat{\Delta}_0\right)
h(z,w;t)=0,
\ee
with an initial condition
\be
\lim_{t\to 0}h(z,w;t) 
 = \delta^2 (z-w).
\nn
\ee
The Laplacian $\hat{\Delta}_0$ is defined on the curved Riemann surface
and includes the spin connections, but if we expand the Laplacian 
around the flat-space Laplacian
\be
\hat{\Delta}_0 = -4\del_z\del_\zb + \hat{V}_0,
\nn
\ee
and treat $\hat{V}_0$ as a perturbation,
the leading part of the heat kernel is solved to yield
\be
h(z,w;t) = \frac{1}{4\pi t} e^{-|z-w|^2/4t}+\cdots.
\nn
\ee

Thus we find
\be
\begin{split}
\tr \frac{1}{(  \hat{\Delta}_0+ |\phi_0|^2  )^2}{\cal O}
&= \frac{1}{4\pi} \int_0^\infty dt \, e^{-t|\phi_0|^2}\int d^2 z \,{\cal O}(z) + \cdots\\
&=\frac{1}{4\pi} \frac{1}{|\phi_0|^2} \int d^2 z \,{\cal O}(z) + \cdots,
\end{split}
\ee
taking the limit of the trace (\ref{trace}).
The higher order terms in $1/|\phi_0|^2$ contain the higher pole of $\bar{\phi}_0$.
We only need the above leading term since
these higher poles would disappear after the integration of $\bar{\phi}_0$.


\end{document}